\documentclass[aps,superscriptaddress,10pt,nobibnotes,notitlepage,twocolumn,longbibliography]{revtex4-1}

\usepackage[normalem]{ulem}

\usepackage[latin1]{inputenc}
\usepackage{graphicx}
\usepackage{float}
\usepackage{latexsym}
\usepackage{graphicx}
\usepackage{amssymb}
\usepackage{amsmath}
\usepackage{amsfonts}
\usepackage[colorlinks=true,citecolor=blue,hyperfootnotes=false]{hyperref}

\usepackage{cool}
\usepackage{dcolumn}
\usepackage{textcomp}
\usepackage[dvipsnames]{xcolor}
\usepackage{xfrac}
\usepackage{slashed}
\usepackage{multirow}

\def\be{\begin{equation}}
\def\ee{\end{equation}}
\def\figs/B{B}
\def\bea{\begin{eqnarray}}
\def\eea{\end{eqnarray}}
\def\bg{\begin{eqnarray}}
\def\nd{\end{eqnarray}}
\def\sin{{\rm sin}}
\def\cos{{\rm cos}}

\def\ln{{\rm log}}

\def\beq{\begin{equation}}
\def\eeq{\end{equation}}

\begin{document}

\title{Primordial Black Holes from Multifield Inflation with Nonminimal Couplings }

\author{Sarah R.~Geller}

\affiliation{Department of Physics, Massachusetts Institute of Technology, Cambridge, MA 02139, USA}

\author{Wenzer Qin}

\affiliation{Department of Physics, Massachusetts Institute of Technology, Cambridge, MA 02139, USA}

\author{Evan McDonough}

\affiliation{Department of Physics, University of Winnipeg, Winnipeg MB, R3B 2E9, Canada}

\author{David I.~Kaiser}
\affiliation{Department of Physics, Massachusetts Institute of Technology, Cambridge, MA 02139, USA}
\email{dikaiser@mit.edu}

\begin{abstract}
Primordial black holes (PBHs) provide an exciting prospect for accounting for dark matter. In this paper, we consider inflationary models that incorporate realistic features from high-energy physics---including multiple interacting scalar fields and nonminimal couplings to the spacetime Ricci scalar---that could produce PBHs with masses in the range required to address the present-day dark matter abundance. Such models are consistent with supersymmetric constructions, and only incorporate operators in the effective action that would be expected from generic effective field theory considerations.  The models feature potentials with smooth large-field plateaus together with small-field features that can induce a brief phase of ultra-slow-roll evolution. Inflationary dynamics within this family of models yield predictions for observables in close agreement with recent measurements, such as the spectral index of primordial curvature perturbations and the ratio of power spectra for tensor to scalar perturbations. As in previous studies of PBH formation resulting from a period of ultra-slow-roll inflation, we find that at least one dimensionless parameter must be highly fine-tuned to produce PBHs in the relevant mass-range for dark matter. Nonetheless, we find that the models described here yield accurate predictions for a significant number of observable quantities using a smaller number of relevant free parameters.
\end{abstract}

\date{\today}

\maketitle

%\color{black}{\tableofcontents}

\section{Introduction}

Primordial black holes (PBHs) were first postulated more than half a century ago \cite{ZeldovichNovikov1967,Hawking1971,CarrHawking1974}, and they remain a fascinating theoretical curiosity. In recent years, many researchers have realized that PBHs provide an exciting prospect for accounting for dark matter. Rather than requiring some as-yet unknown elementary particles beyond the Standard Model, dark matter might consist of a large population of PBHs that formed very early in cosmic history. See Refs.~\cite{Carr:2020xqk,Green:2020jor,Villanueva-Domingo:2021spv} for recent reviews.

Much activity has focused on mechanisms by which PBHs could form from density perturbations that were generated during early-universe inflation. When overdensities with magnitude above some critical threshold re-enter the Hubble radius after the end of inflation, they induce gravitational collapse into black holes. Many studies have focused on specific inflationary models that can yield appropriate perturbations; PBH formation following hybrid inflation has garnered particular attention \cite{Garcia-Bellido:1996mdl,Lyth:2010zq,Bugaev:2011wy,Halpern:2014mca,Clesse:2015wea,Kawasaki:2015ppx}. Others have found clever ways to engineer
desired features of a given model so as to generate PBHs, by inserting specific features into the potential and/or non-canonical kinetic terms for the field(s) driving inflation. See, e.g., Refs.~\cite{Garcia-Bellido:2017mdw,Ezquiaga:2017fvi,Kannike:2017bxn,Germani:2017bcs,Motohashi:2017kbs,Di:2017ndc,Ballesteros:2017fsr,Pattison:2017mbe,Passaglia:2018ixg,Biagetti:2018pjj,Byrnes:2018txb,Carrilho:2019oqg,Ashoorioon:2019xqc,Aldabergenov:2020bpt,Ashoorioon:2020hln,Inomata:2021uqj,Inomata:2021tpx,Pattison:2021oen,Lin:2020goi,Palma:2020ejf,Yi:2020cut,Iacconi:2021ltm,Kallosh:2022vha,Ashoorioon:2022raz,Frolovsky:2022ewg,Aldabergenov:2022rfc}.

In this work we explore possibilities for the production of PBHs within well-motivated models of inflation that feature realistic ingredients from high-energy theory. In particular, we consider models with several interacting scalar fields, each of which includes a nonminimal coupling to the spacetime Ricci scalar. This family of models includes---but is more general than---well-known models such as Higgs inflation \cite{Bezrukov:2007ep} and $\alpha$-attractor models \cite{Kallosh:2013maa,Kallosh:2013yoa,Galante:2014ifa}. For example, the Higgs sector of the Standard Model includes four scalar degrees of freedom, all of which remain in the spectrum at high energies within renormalizable gauges \cite{Mooij:2011fi,Greenwood:2012aj}. Moreover, every candidate for Beyond Standard Model physics includes even more scalar degrees of freedom at high energies \cite{Lyth:1998xn,Mazumdar:2010sa}. Likewise, nonminimal couplings in the action of the form $\xi \phi^2 R$, where $\phi$ is a scalar field, $R$ is the spacetime Ricci scalar, and $\xi$ a dimensionless constant, are required for renormalization and, more generally, are induced by quantum corrections at one-loop order even if the couplings $\xi$ vanish at tree-level \cite{Callan:1970ze,Bunch:1980br,Bunch:1980bs,Birrell:1982ix,Odintsov:1990mt,Buchbinder:1992rb,Faraoni:2000gx,Parker:2009uva,Markkanen:2013nwa,Kaiser:2015usz}. The couplings $\xi$ generically increase with energy scale under renormalization-group flow with no UV fixed point \cite{Odintsov:1990mt,Buchbinder:1992rb}, and hence they can be large $(\vert \xi \vert \gg 1$) at the energy scales relevant for inflation. Finally, although the models we study need not make recourse to supersymmetry or supergravity, we find they can be realized in simple supergravity setups, including in models that simultaneously realize the observed cosmological constant. 

Inflationary dynamics in the family of models we consider generically yield predictions for observable quantities, such as the spectral index of primordial curvature perturbations and the ratio of power spectra for tensor and scalar perturbations, in close agreement with recent measurements \cite{Kaiser:2012ak,Kaiser:2013sna,Schutz:2013fua}. Such models also generically yield efficient post-inflation reheating, typically producing a radiation-dominated equation of state and a thermal spectrum of decay products within $N_{\rm reh} \sim {\cal O} (1)$ e-folds after the end of inflation \cite{Bezrukov:2008ut,Garcia-Bellido:2008ycs,Child:2013ria,DeCross:2015uza,DeCross:2016fdz,DeCross:2016cbs,Figueroa:2016dsc,Repond:2016sol,Ema:2016dny,Sfakianakis:2018lzf,Rubio:2019ypq,Nguyen:2019kbm,vandeVis:2020qcp,Iarygina:2020dwe,Ema:2021xhq,Figueroa:2021iwm,Dux:2022kuk}. Hence such models represent an important class in which to consider PBH production.

We find that such models provide a natural framework within which PBHs could form. As in previous studies that focused on the formation of PBHs from a phase of ultra-slow-roll inflation \cite{Garcia-Bellido:2017mdw,Ezquiaga:2017fvi,Kannike:2017bxn,Germani:2017bcs,Motohashi:2017kbs,Di:2017ndc,Ballesteros:2017fsr,Pattison:2017mbe,Passaglia:2018ixg,Byrnes:2018txb,Biagetti:2018pjj,Carrilho:2019oqg,Inomata:2021tpx,Inomata:2021uqj,Pattison:2021oen}, we also find that to produce perturbation spectra relevant for realistic PBH scenarios, at least one dimensionless parameter must be highly fine-tuned. Nonetheless, we find that such models can yield accurate predictions for a significant number of observable quantities using a smaller number of relevant free parameters. In this paper we focus on the general mechanisms by which such models can produce PBHs, and defer to later work a more thorough analysis of the full parameter space.

In Section \ref{sec:model} we introduce the family of multifield models on which we focus and identify generic features of their dynamics. Section \ref{sec:PBHform} considers the formation of PBHs after the end of inflation, including how the production of PBHs is affected by changes to various model parameters. Concluding remarks follow in Section \ref{sec:discussion}. In Appendix \ref{appPerturbations}, we review important features of gauge-invariant perturbations in multifield models, while in Appendix \ref{appSUGRA} we demonstrate how this family of models can be realized within a supergravity framework. Appendix \ref{appTheta} includes additional details about our analytic solution for the fields' trajectory through field space during inflation.
Throughout this paper we adopt ``natural units" ($c = \hbar = k_B = 1$) and work in terms of the reduced Planck mass, $M_{\rm pl} \equiv 1 / \sqrt{ 8 \pi G} = 2.43 \times 10^{18} \, {\rm GeV}$.

\section{Multifield Model and Dynamics}
\label{sec:model}

\subsection{Multifield Formalism}
\label{sec:multifieldformalism}

We begin with a brief review of multifield dynamics for background quantities and linearized fluctuations, following the notation of Ref.~\cite{Kaiser:2012ak}. See also Appendix \ref{appPerturbations}, Refs.~\cite{Sasaki:1995aw,Langlois:2008mn,Peterson:2010np,Gong:2011uw}, and Ref.~\cite{Gong:2016qmq} for a review of gauge-invariant perturbations in multifield models. We consider models with ${\cal N}$ scalar fields $\phi^I (x^\mu)$ with $I = 1, 2, ... , \, {\cal N}$, and work in $(3 + 1)$ spacetime dimensions. In the Jordan frame, the action may be written
%%%%%%
\beq
\tilde{S} = \int d^4 x \sqrt{-\tilde{g}} \left[ f (\phi^I ) \tilde{R}  - \frac{1}{2} \delta_{IJ} \tilde{g}^{\mu\nu}  \partial_\mu \phi^I \partial_\nu \phi^J - \tilde{V} (\phi^I) \right] ,
\label{SJ}
\eeq
where $f (\phi^I)$ denotes the fields' nonminimal couplings and tildes indicate quantities in the Jordan frame. After performing a conformal transformation by rescaling $\tilde{g}_{\mu\nu} (x) \rightarrow g_{\mu\nu} (x) = \Omega^2 (x) \tilde{g}_{\mu\nu} (x)$ with conformal factor
%%%%%
\beq
\Omega^2 (x) = \frac{2}{M_{\rm pl}^2} f (\phi^I (x) ) ,
\label{Omega}
\eeq
we may write the action in the Einstein frame as \cite{Kaiser:2010ps}
%%%%%%
\beq
S = \int d^4 x \sqrt{-g} \left[ \frac{ M_{\rm pl}^2}{2} R - \frac{1}{2} {\cal G}_{IJ} g^{\mu\nu} \partial_\mu \phi^I \partial_\nu \phi^J - V (\phi^I) \right] ,
\label{SE}
\eeq
where the potential in the Einstein frame is stretched by the conformal factor,
%%%%%
\beq
V (\phi^I) = \frac{ M_{\rm pl}^4}{4 f^2 (\phi^I) } \tilde{V} (\phi^I ) .
\label{VEconformal}
\eeq
The nonminimal couplings induce a curved field-space manifold in the Einstein frame with associated field-space metric
%%%%%
\beq
{\cal G}_{IJ} (\phi^K) = \frac{M_{\rm pl}^2}{2 f (\phi^K)} \left[ \delta_{IJ} + \frac{ 3}{f (\phi^K)} f_{, I} f_{, J} \right] ,
\label{GIJgeneral}
\eeq
where $f_{, I} \equiv \partial f / \partial \phi^I$. For ${\cal N} \geq 2$ fields with nonminimal couplings, one cannot canonically normalize all of the fields while retaining the Einstein-Hilbert form of the gravitational part of the action \cite{Kaiser:2010ps}.

We consider perturbations around a spatially flat Friedmann-Lema\^{i}tre-Robertson-Walker (FLRW) line element, as discussed further in Appendix \ref{appPerturbations}, and separate each scalar field into a spatially homogeneous vacuum expectation value and spatially varying fluctuations:
%%%%
\beq
\phi^I (x^\mu) = \varphi^I (t) + \delta \phi^I (x^\mu).
\label{phivarphi}
\eeq
The equation of motion for the spatially homogeneous background fields then takes the form
%%%%
\beq
{\cal D}_t \dot{\varphi}^I + 3 H \dot{\varphi}^I + {\cal G}^{IK} V_{, K} = 0 ,
\label{eomvarphi}
\eeq
where $H \equiv \dot{a} / a$ and ${\cal D}_t A^I = \dot{\varphi}^J {\cal D}_J A^I$ for any field-space vector $A^I$, and where the covariant derivative  ${\cal D}_J$ 
employs the usual Levi-Civita connection associated with the metric ${\cal G}_{IJ}$. Since we consider only linearized fluctuations in this paper, we may set ${\cal G}_{IJ} (\phi^K) \rightarrow {\cal G}_{IJ} (\varphi^K)$, so that components of the field-space metric depend only on time. The magnitude of the background fields' velocity vector is given by
%%%%
\beq
\vert \dot{\varphi}^I \vert \equiv \dot{\sigma} = \sqrt{ {\cal G}_{IJ} \, \dot{\varphi}^I \dot{\varphi}^J } ,
\label{dotsigmadef}
\eeq
in terms of which we may write the unit vector
%%%%
%%%%%%%%
\beq
\hat{\sigma}^I \equiv \frac{ \dot{\varphi}^I } {\dot{\sigma}} 
\label{hatsigma}
\eeq
which points along the background fields' direction of motion in field space. The quantity
%%%%
\beq
\hat{s}^{IJ} \equiv {\cal G}^{IJ} - \hat{\sigma}^I \hat{\sigma}^J
\label{sIJ}
\eeq
projects onto the subspace of the field-space manifold perpendicular to the background fields' motion.

In terms of $\dot{\sigma}$, the equations of motion for background quantities may be written \cite{Kaiser:2012ak}
%%%%%
\beq
\begin{split}
\ddot{\sigma} &+ 3 H \dot{\sigma} + V_{, \sigma} = 0 , \\
H^2 &= \frac{1}{ 3 M_{\rm pl}^2} \left[ \frac{1}{2} \dot{\sigma}^2 + V \right] , \\
\dot{H} &= - \frac{1}{ 2 M_{\rm pl}^2} \dot{\sigma}^2 ,
\end{split}
\label{eombackground}
\eeq
where
%%%%%
\beq
V_{, \sigma} \equiv \hat{\sigma}^I V_{, I} .
\label{Vsigma}
\eeq
The covariant turn-rate vector is defined as \cite{Kaiser:2012ak}
%%%%
\beq
\omega^I \equiv {\cal D}_t \hat{\sigma}^I = - \frac{1}{ \dot{\sigma}} V_{, K} \hat{s}^{IK} ,
\label{omegadef}
\eeq
where the last expression follows upon using Eqs.~(\ref{eomvarphi}), (\ref{sIJ}), and (\ref{eombackground}). The usual slow-roll parameter takes the form
%%%%%
\beq
\epsilon \equiv - \frac{ \dot{H}}{H^2} = \frac{ 1}{2 M_{\rm pl}^2} \frac{ \dot{\sigma}^2}{H^2} ,
\label{epsilon}
\eeq
where the last expression follows upon using Eq.~(\ref{eombackground}). We define the end of inflation $t_{\rm end}$ via $\epsilon (t_{\rm end}) = 1$, which corresponds to $\ddot{a} (t_{\rm end}) = 0$, the end of accelerated expansion.

In addition to $\epsilon$, we consider a second slow-roll parameter
%%%%%
\beq
\eta \equiv 2 \epsilon - \frac{ \dot{\epsilon}}{2 H \epsilon}.
\label{etadef}
\eeq
Using Eqs.~(\ref{eombackground}) and (\ref{epsilon}) we see that, in general,
%%%%
\beq
\frac{\dot{\epsilon}}{2 H \epsilon} = \frac{\ddot{\sigma}}{H \dot{\sigma}} + \epsilon .
\label{ddotsigmaeta}
\eeq
During ordinary slow-roll evolution $\vert \ddot{\sigma} \vert \ll \vert 3 H \dot{\sigma} \vert$, and the top line of Eq.~(\ref{eombackground}) becomes $3 H \dot{\sigma} \simeq - V_{, \sigma}$. Under those conditions $\eta \sim \epsilon < 1$. However, during so-called ultra-slow-roll, the potential becomes nearly flat, $V_{, \sigma} \simeq 0$, and hence the equation of motion for the background fields becomes $\ddot{\sigma} \simeq - 3 H \dot{\sigma}$. In that case, $\epsilon$ becomes exponentially smaller than 1 and 
%%%%%
\beq
\eta \rightarrow 3 \quad\quad (\textrm{ultra-slow-roll}). 
\label{etaUSR}
\eeq
Eq.~(\ref{etadef}) then yields $\dot{\epsilon} + 6 H \epsilon \simeq 0$. Given $H \simeq {\rm constant}$ during ultra-slow-roll evolution (consistent with $\epsilon \ll 1$), the kinetic energy density of the background fields $\rho_{\rm kin} = \dot{\sigma}^2 / 2 = M_{\rm pl}^2 H^2 \epsilon$ rapidly redshifts as $\rho_{\rm kin} (t) \sim a^{-6} (t)$ \cite{Kinney:2005vj,Martin:2012pe,Namjoo:2012aa,Romano:2015vxz,Germani:2017bcs,Dimopoulos:2017ged,Biagetti:2018pjj,Byrnes:2018txb,Pattison:2018bct,Carrilho:2019oqg,Inomata:2021tpx,Inomata:2021uqj,Pattison:2017mbe,Pattison:2019hef,Pattison:2021oen}.

The gauge-invariant Mukhanov-Sasaki variables $Q^I$ are constructed as linear combinations of metric perturbations and the field fluctuations, as in Eq.~(\ref{QIdef}). We may project the perturbations $Q^I$ into adiabatic ($Q_\sigma$) and isocurvature ($\delta s^I$) components \cite{Gordon:2000hv,Wands:2002bn,Bassett:2005xm,Kaiser:2012ak},
%%%%
\beq
Q^I = \hat{\sigma}^I Q_\sigma + \delta s^I ,
\label{QIadiso}
\eeq
where
%%%%%
\beq
Q_\sigma \equiv \hat{\sigma}_J Q^J , \>\> \delta s^I \equiv \hat{s}^I_{\>\> J} Q^J .
\label{Qsigmadeltasdef}
\eeq
For two-field models, as we consider below, the isocurvature perturbations are characterized by a field-space scalar $Q_s$ defined via \cite{McDonough:2020gmn}
%%%%%
\beq
\delta s^J = \epsilon^{IJ} \hat{\sigma}_I Q_s ,
\label{deltasQs}
\eeq
where $\epsilon^{IJ} \equiv [ {\rm det} ({\cal G}_{IJ} ) ]^{-1/2} \, \bar{\epsilon}^{IJ}$ and $\bar{\epsilon}^{IJ}$ is the usual antisymmetric Levi-Civita symbol. The equations of motion for Fourier modes of comoving $k$, $Q_\sigma (k, t)$ and $Q_s (k, t)$, are given in Eqs.~(\ref{Qsigmaeom})--(\ref{Qseom}), from which it is clear that the adiabatic and isocurvature perturbations decouple for non-turning trajectories, for which $\vert \omega^I \vert = 0$. In addition, the amplitude of isocurvature perturbations will be suppressed as $Q_s (k, t) \sim a^{-3/2} (t)$ while $\mu_s^2 / H^2 \gg 1$, where the mass of the isocurvature perturbations, $\mu_s^2$, is given in Eq.~(\ref{mus}). Hence if $\omega^2 \ll H^2$ or $\mu_s^2 / H^2 \gg 1$, or both, there will be negligible transfer of power from the isocurvature to the adiabatic modes \cite{Gordon:2000hv,Wands:2002bn,Bassett:2005xm,Kaiser:2012ak,Kaiser:2013sna,Kaiser:2015usz,Schutz:2013fua,Langlois:2008mn,Peterson:2010np,Gong:2011uw,Gong:2016qmq,McDonough:2020gmn}.

The adiabatic perturbation is proportional to the gauge-invariant curvature perturbation \cite{Kaiser:2012ak}
%%%%%%
\beq
{\cal R} = \frac{ H}{ \dot{\sigma}} Q_\sigma = \frac{ Q_\sigma}{M_{\rm pl} \sqrt{ 2 \epsilon} } ,
\label{Rdef}
\eeq
where the last equality 
%expression 
follows from Eq.~(\ref{epsilon}). To avoid confusion, we adopt the convention of 
Ref.~\cite{Iacconi:2021ltm} and denote the curvature perturbation as ${\cal R}$ and the Ricci scalar of the field-space manifold as ${\cal R}_{\rm fs}$. The dimensionless power spectrum for the curvature perturbations is defined as 
usual:
%%%%%
\beq
{\cal P}_{\cal R} (k) \equiv \frac{ k^3}{ 2 \pi^2} \vert {\cal R}_k  \vert^2 .
\label{PRdef}
\eeq
Given the form of Eqs.~(\ref{Rdef})--(\ref{PRdef}), there are at least two distinct mechanisms by which inflationary dynamics could yield a large spike in ${\cal P}_{\cal R} (k)$ at relevant scales $k$, which could produce PBHs after inflation: either by amplifying $Q_\sigma (k,t)$ or by reducing $\epsilon (t)$. The former could occur by some feature of the dynamics such as a brief tachyonic phase for certain modes $k$, akin to what occurs in hybrid inflation models at the waterfall transition \cite{Garcia-Bellido:1996mdl,Lyth:2010zq,Bugaev:2011wy,Halpern:2014mca,Clesse:2015wea,Kawasaki:2015ppx}, or by a transfer of power from isocurvature to adiabatic modes during a fast turn in field space \cite{Palma:2020ejf,McDonough:2020gmn,Braglia:2020eai,Braglia:2020fms,Iacconi:2021ltm,Kallosh:2022vha}. The other typical mechanism---by which the slow-roll parameter $\epsilon$ falls by several orders of magnitude, $0 \leq \epsilon \ll 1$---occurs during ultra-slow-roll evolution \cite{Garcia-Bellido:2017mdw,Ezquiaga:2017fvi,Kannike:2017bxn,Germani:2017bcs,Motohashi:2017kbs,Di:2017ndc,Ballesteros:2017fsr,Pattison:2017mbe,Passaglia:2018ixg,Byrnes:2018txb,Biagetti:2018pjj,Carrilho:2019oqg,Pattison:2021oen}, which can occur even if there is no turning of the fields' trajectory in field-space. A related but distinct mechanism involves particle production as the inflaton crosses a step-like feature in the potential, followed by ultra-slow-roll evolution to amplify the perturbations associated with the produced particles \cite{Inomata:2021tpx,Inomata:2021uqj}.

The models on which we focus here generically include periods of ultra-slow-roll evolution near the end of inflation. In order for such an ultra-slow-roll phase to produce a large spike in ${\cal P}_{\cal R} (k)$, quantum fluctuations of the fields must not whisk the system past the region of the potential in which $V_{, \sigma} \simeq 0$ too quickly, or else inflation will end before significant amplification of ${\cal P}_{\cal R} (k)$ can occur \cite{Germani:2017bcs,Motohashi:2017kbs,Di:2017ndc,Ballesteros:2017fsr,Dimopoulos:2017ged,Pattison:2017mbe,Biagetti:2018pjj,Byrnes:2018txb,Pattison:2019hef,Carrilho:2019oqg,Passaglia:2018ixg,Pattison:2021oen,Inomata:2021uqj,Inomata:2021tpx}. Backreaction from quantum fluctuations 
yields a variance of the kinetic energy density for the system \cite{Inomata:2021tpx}
%%%%%
\beq
\langle (\Delta K )^2 \rangle \simeq \frac{ 3 H^4}{4 \pi^2} \rho_{\rm kin} ,
\label{varianceK}
\eeq
where $\rho_{\rm kin} = \dot{\sigma}^2 / 2$ is the background fields' unperturbed kinetic energy density. Classical evolution will dominate quantum diffusion during ultra-slow-roll evolution if $\rho_{\rm kin} > \sqrt{\langle (\Delta K )^2 \rangle}$. Upon using Eq.~(\ref{epsilon}), this criterion becomes
%%%%%
\beq
\epsilon_{\rm usr} > \frac{3}{ 4 \pi^2} \left( \frac{ H}{ M_{\rm pl} } \right)^2 .
\label{epsUSRdominate}
\eeq
Comparing with Eq.~(\ref{PRHepsilon}), we see that Eq.~(\ref{epsUSRdominate}) is equivalent to ${\cal P}_{\cal R} (k) < 1/6$ \cite{Inomata:2021tpx}. Within the regions of parameter space that we consider in Sections \ref{sec:trajectories} and \ref{sec:PBHsUSR}, the criterion of Eq.~(\ref{epsUSRdominate}) is always satisfied, such that during ultra-slow-roll, classical evolution of the background fields continues to dominate over quantum diffusion, allowing for a robust amplification of curvature perturbations. 

In the absence of a transfer of power from isocurvature to adiabatic perturbations, predictions for observables relevant to the cosmic microwave background radiation (CMB) revert to the familiar and effectively single-field forms \cite{Kaiser:2012ak,Kaiser:2013sna}. Explicit expressions for the spectral index $n_s (k_*)$, the running of the spectral index $\alpha (k_*) \equiv (d n_s (k_*) / d {\rm ln} k)\vert_{k_*}$, and the tensor-to-scalar ratio $r (k_*)$ may be found in Eqs.~(\ref{nsdef})--(\ref{rTtoS}); here $k_* = 0.05 \, {\rm Mpc}^{-1}$ is the comoving CMB pivot scale. Likewise, inherently multifield features, such as the fraction of primordial isocurvature perturbations $\beta_{\rm iso} (k_*, t_{\rm end})$, which is defined in Eq.~(\ref{betaisodef}), and primordial non-Gaussianity $f_{\rm NL}$, defined in Eq.~(\ref{fNLSFA2}), generically remain small for multifield models in which the isocurvature modes remain heavy throughout inflation ($\mu_s^2 \gg H^2$) and the turn-rate remains negligible ($\omega^2 \ll H^2$) \cite{Kaiser:2012ak,Kaiser:2013sna,Schutz:2013fua,Kaiser:2015usz,Gordon:2000hv,Langlois:2008mn,DiMarco:2002eb,Bernardeau:2002jy,Seery:2005gb,Yokoyama:2007dw,Byrnes:2008wi,Peterson:2010mv,Chen:2010xka,Byrnes:2010em,Gong:2011cd,Elliston:2011dr,Elliston:2012ab,Seery:2012vj,Mazumdar:2012jj,Wands:2002bn,Bassett:2005xm,Peterson:2010np,Gong:2011uw,Gong:2016qmq,McDonough:2020gmn}.

\subsection{Supersymmetric Two-Field Models}
\label{sec:susymodels}

For the remainder of this paper we consider supersymmetric two-field models, in which  supersymmetry is spontaneously broken. These models naturally arise in both global supersymmetry and supergravity.
Although our framework does not depend strongly on supersymmetric motivations, the supersymmetric framework provides a codex for translating a relatively large number of effective field theory parameters to a much smaller set of parameters that govern the UV completion in supergravity, which is valid at least at tree-level. The desired nonminimal couplings can then be realized in a manifestly supersymmetric manner, e.g., as in Refs.~\cite{Kallosh:2010ug,Kallosh:2013hoa}, in the superconformal approach to supergravity \cite{Freedman:2012zz}, or else generated via quantum effects once supersymmetry has been spontaneously broken. Here we provide a brief overview. Additional details may be found in Appendix \ref{appSUGRA} and Ref.~\cite{Freedman:2012zz}.

As mentioned, at the energy scales relevant for inflation, the construction yields specific arrangements among various dimensionless coupling constants, but the field operators that appear in the action include only generic dimension-4 operators that should be included in {\it any} self-consistent effective field theory for two interacting scalar fields in $(3 + 1)$ spacetime dimensions. This sort of SUSY pattern imprinted on low-energy physics has been discussed 
in the context of CMB non-Gaussianity from supersymmetric higher-spin fields \cite{Alexander:2019vtb}. 

We focus on inflation models that may be realized in the global supersymmetry limit of supergravity. The model is specified by a K\"{a}hler potential $\tilde{K}$ and superpotential $\tilde{W}$ in the Jordan frame, given by
\begin{equation}
    \tilde{K} (\Phi, \bar{\Phi} ) = - \frac{1}{2} \displaystyle \sum _{I=1} ^2 ( \Phi^I - \bar{\Phi}^{\bar{I}})^2
    \label{KahlerJordan1}
\end{equation}
and
\begin{equation}
    \tilde{W} (\Phi) =  \sqrt{2} \, \mu b_{IJ} \Phi^I \Phi^J + 2 c_{IJK} \Phi^I \Phi^J \Phi^K ,
    \label{Wtilde1}
\end{equation}
with indices $I, J, K \in \{ 1, 2 \}$. We select $\tilde{K}$ so as to provide canonical kinetic terms for the real and imaginary components of the scalar fields $\varpi^I$ associated with each chiral superfield $\Phi^I$ (as further discussed in Appendix \ref{appSUGRA}), and insert factors of $\sqrt{2}$ and $2$ in the superpotential $\tilde{W}$ to reduce clutter in the resulting equations. The coefficients $b_{IJ}$ and $c_{IJK}$ in $\tilde{W}$ are real-valued dimensionless coefficients, and repeated indices are trivially summed over. We omit possible constant and linear contributions to $\tilde{W}$, since non-renormalization of $\tilde{W}$ \cite{Grisaru:1979wc,Seiberg:1993vc} provides the freedom to do so. Expanding Eq.~(\ref{Wtilde1}), we may express $\tilde{W}$ as
\begin{equation}
    \begin{split}
    \tilde{W} &= \sqrt{2} \, b_1 \mu (\Phi_1 )^2 + \sqrt{2}  \, b_2 \mu (\Phi_2 )^2 +  2 c_1 (\Phi_1)^3  \\
   & \quad + 2 c_2 (\Phi_1)^2 \Phi_2 + 2 c_3 \Phi_1 (\Phi_2)^2 + 2 c_4 (\Phi_2 )^3 ,
    \label{Wtilde}
\end{split}
\end{equation}
where we have defined $b_1 \equiv b_{11}$, $b_2 \equiv b_{22}$, $c_1 \equiv c_{111}$, $c_2 \equiv (c_{112}+c_{121}+c_{211})$, $c_3 \equiv (c_{122}+c_{212}+c_{221})$, and $c_4 \equiv c_{222}$. We set the coupling $b_{12}$ for the quadratic cross-term $\mu \Phi_1 \Phi_2$ to zero for simplicity but without loss of generality, since this choice merely amounts to a choice of coordinates on field space.

The K\"{a}hler potential and superpotential together determine the scalar potential as
\begin{equation}
    \tilde{V} = e^{ \tilde{K}/M_{\rm pl}^2}\left( |D \tilde{W}|^2 - 3M_{\rm pl}^{-2}|\tilde{W}|^2  \right) ,
    \label{VtildeSUSY}
\end{equation}
where $D_I \equiv \partial_I + M_{\rm pl}^{-2} \tilde{K}_{, I}$ denotes a K\"{a}hler covariant derivative \cite{Freedman:2012zz}. The explicit tilde on $V$ indicates that the chiral superfields $\Phi^I$ are assumed to be nonminimally coupled to gravity, either through a manifestly supersymmetric setup or through quantum effects below the SUSY breaking scale, making the  expression for $\tilde{V}$ in Eq.~(\ref{VtildeSUSY}) the Jordan-frame potential. 

The choice of K\"{a}hler potential in Eq.~(\ref{KahlerJordan1}) guarantees that the imaginary parts of the scalar components of $\Phi^I$ are heavy during inflation, $m^2_{\psi} > H^2$, where $\Phi^I = \varpi^I + ...$ for complex scalar fields $\varpi^I$, and $\varpi^I = ( \phi^I + i \psi^I) / \sqrt{2}$, with $\phi^I$ and $\psi^I$ real-valued scalar fields. In the global supersymmetry limit ($\vert \Phi^I \vert^2 / M^2_{\rm pl}\rightarrow \infty$), the scalar potential can then be expressed as simply
\beq
\tilde{V} (\phi, \chi) \simeq \sum_I \left\vert \frac{ \partial W }{\partial \Phi^I } \right\vert^2_{\Phi^I \rightarrow \varpi^I} ,
\label{Vtildedef}
\eeq
where we label the real-valued scalar components of the chiral superfields as $\Phi_1 = \phi/\sqrt{2}$ and $\Phi_2 = \chi/\sqrt{2}$. We discuss additional details of the embedding in supergravity in Appendix \ref{appSUGRA}.

\subsection{The Einstein-Frame Scalar Potential}

The full form of $\tilde{V} (\phi, \chi)$ appears in Appendix \ref{appSUGRA}. For our two-field models, it is convenient to adopt polar coordinates for the field-space manifold,
%%%%%
\beq
\phi (t) = r (t) \, \cos \theta (t) , \>\> \chi (t) = r (t) \, \sin \theta (t) ,
\label{phichirtheta}
\eeq
with $r \geq 0$ and $0 \leq \theta \leq 2 \pi$. Then the Jordan-frame scalar potential of Eq.~(\ref{Vtildedef}) takes the form
%%%%%
\beq
\tilde{V} (r, \theta) = {\cal B} (\theta) \mu^2 r^2 + {\cal C} (\theta) \mu r^3 + {\cal D} (\theta) r^4
\label{Vtildertheta}
\eeq
with
%%%%%
\beq
\begin{split}
    {\cal B} (\theta) &\equiv 4 b_1^2 \cos^2 \theta + 4 b_2^2 \sin^2 \theta , \\
    {\cal C} (\theta) &\equiv 12 b_1 c_1 \cos^3 \theta + 4  (2b_1 + b_2) c_2 \cos^2 \theta \sin \theta \\
    &\quad + 4 (b_1 + 2 b_2) c_3 \cos \theta \sin^2 \theta + 12 b_2 c_4 \sin^3 \theta , \\
    {\cal D} (\theta) &\equiv (9 c_1^2 + c_2^2) \cos^4 \theta + 4 c_2 (3 c_1 + c_3) \cos^3 \theta \sin\theta \\
    &\quad + (4 c_2^2 + 6 c_1 c_3 + 6 c_2 c_4 + 4 c_3^2) \cos^2 \theta \sin^2\theta \\
    &\quad+ 4 c_3 (c_2 + 3 c_4) \cos \theta \sin^3 \theta + (9 c_4^2 + c_3^2) \sin^4 \theta .
\end{split}
\label{BCDdef}
\eeq
As mentioned, we consider this scalar potential in conjunction with nonminimal couplings to gravity. In a curved spacetime, scalar fields' self-interactions will generate nonminimal couplings of the form \cite{Callan:1970ze,Bunch:1980br,Bunch:1980bs,Birrell:1982ix,Odintsov:1990mt,Buchbinder:1992rb,Faraoni:2000gx,Parker:2009uva,Markkanen:2013nwa,Kaiser:2015usz}
%%%%%
\beq
\begin{split}
f (\phi, \chi) &= \frac{1}{2} \left[ M_{\rm pl}^2 + \xi_\phi \phi^2 + \xi_\chi \chi^2 \right] \\
&= \frac{1}{2} \left[ M_{\rm pl}^2 + r^2 \left( \xi_\phi \cos^2 \theta + \xi_\chi \sin^2 \theta \right) \right] .
\end{split}
\label{frtheta}
\eeq
Hence the action for the scalar degrees of freedom of our models takes the form of Eq.~(\ref{SJ}), with $\tilde{V} (\phi^I)$ given by Eq.~(\ref{Vtildertheta}) and $f (\phi^I)$ by Eq.~(\ref{frtheta}).

\begin{figure*}[t!]
    \centering
    \includegraphics[width=0.43\textwidth]{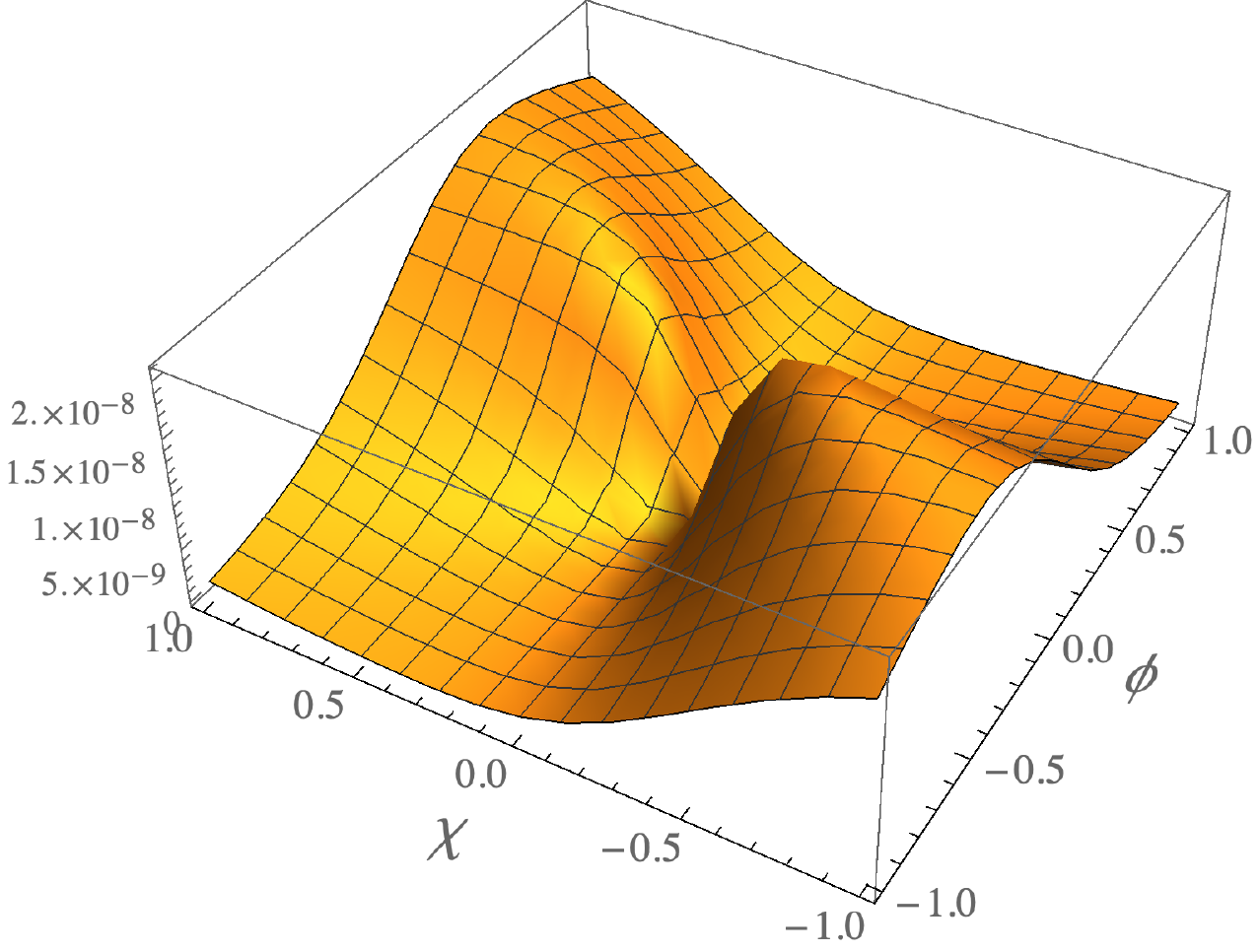} $\quad$ \includegraphics[width=0.43\textwidth]{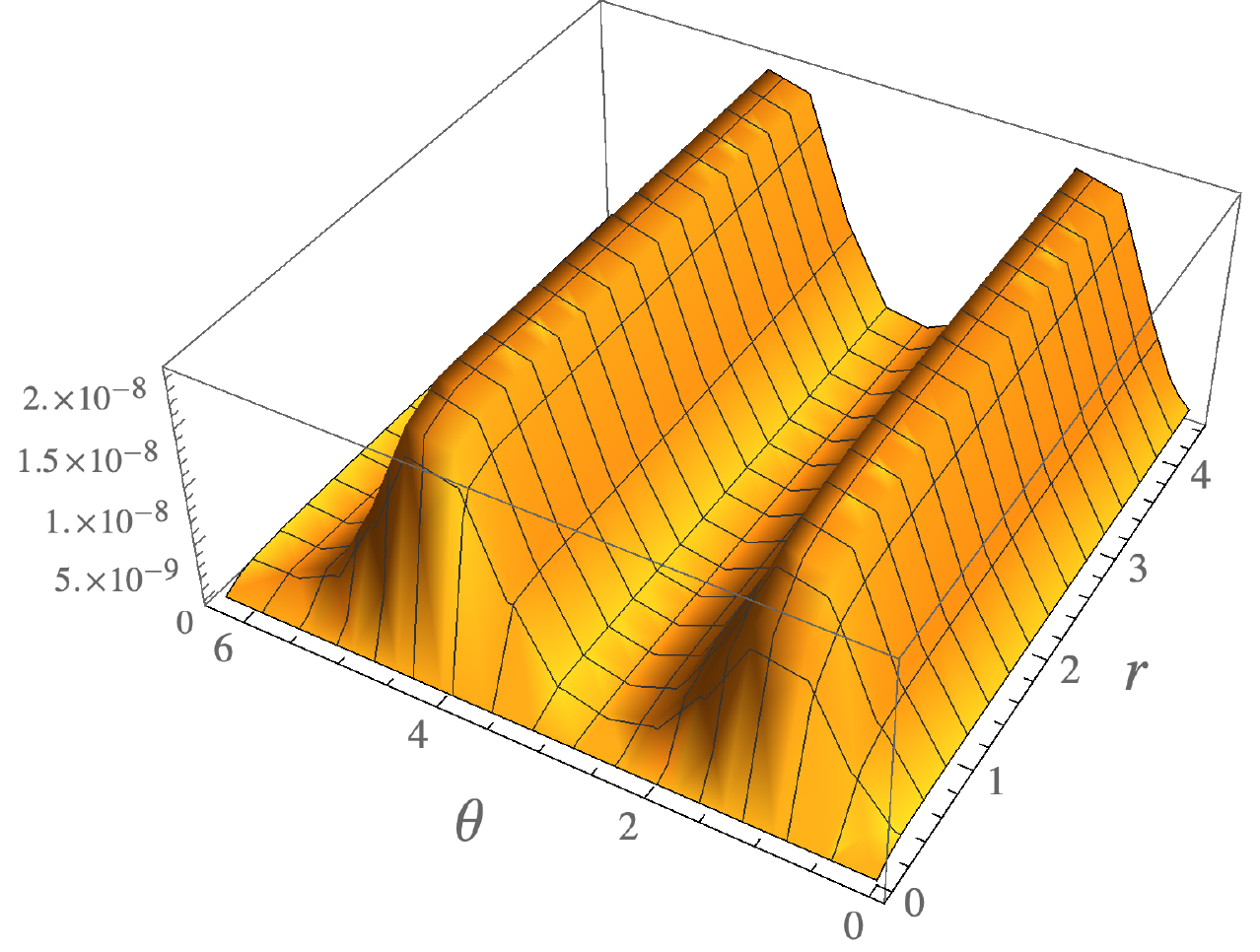}
    \caption{The scalar potential in the Einstein frame, in both $\{ \phi, \chi\}$ ({\it left}) and $\{ r, \theta \}$ ({\it right}) coordinates. Fields are shown in units of $M_{\rm pl}$. The parameters are $\mu = M_{\rm pl}$, $b_1 = b_2 = -1.8 \times 10^{-4}$, $c_1 = 2.5 \times 10^{-4}$, $c_2 = c_3 = 3.57 \times 10^{-3}$, $c_4 = 3.9 \times 10^{-3}$, and $\xi_\phi = \xi_\chi = 100$. 
    }
    \label{fig:VE}
\end{figure*}

Upon transforming to the Einstein frame, the field-space metric ${\cal G}_{IJ}$ in our $\{ r, \theta \}$ coordinates has components
%%%%%%
\beq
\begin{split}
    {\cal G}_{rr} &= \frac{ M_{\rm pl}^2}{2f} \left[ 1 + \frac{3r^2}{f} \left( \xi_\phi \cos^2 \theta + \xi_\chi \sin^2 \theta \right)^2 \right] , \\
    {\cal G}_{r\theta} &= \frac{ M_{\rm pl}^2}{2f} \left( \frac{ 3 r^3}{f} \right) \Big[ \left( \xi_\phi \cos^2\theta + \xi_\chi \sin^2 \theta \right) \\
    &\quad\quad\quad\quad\quad\quad \quad  \times \left( - \xi_\phi + \xi_\chi \right) \cos \theta \sin \theta \Big] , \\
    {\cal G}_{\theta \theta} &= \frac{ M_{\rm pl}^2}{2f} \left[ r^2 + \frac{ 3 r^4}{f} \left( - \xi_\phi + \xi_\chi \right)^2 \cos^2 \theta \sin^2 \theta \right] ,
\label{GIJcomponents}
\end{split}
\eeq
with $f (r, \theta)$ given in Eq.~(\ref{frtheta}). The potential in the Einstein frame becomes
%%%%%
\beq
V (r, \theta) = \frac{ M_{\rm pl}^4}{[ 2f (r, \theta) ]^2} \left[ {\cal B} (\theta) \mu^2 r^2 + {\cal C} (\theta) \mu r^3 + {\cal D} (\theta) r^4 \right] ,
\label{VErtheta}
\eeq
with the coefficients ${\cal B}, {\cal C}$, and ${\cal D}$ given in Eq.~(\ref{BCDdef}). 

The form of $V (\phi^I)$ in Eq.~(\ref{VErtheta}) has a similar structure to the single-field potential studied in Ref.~\cite{Garcia-Bellido:2017mdw}, which included both a cubic self-interaction term and the conformal factor $(M_{\rm pl}^2 + \xi \phi^2)^2$ in the denominator. The potential in Eq.~(\ref{VErtheta}) is also a natural generalization of the two-field models studied in Refs.~\cite{Kaiser:2012ak,Kaiser:2013sna,Schutz:2013fua,Kaiser:2015usz}, for which the numerator included only the term proportional to ${\cal D} (\theta)$. Much as in those multifield studies, the Einstein-frame potential of Eq.~(\ref{VErtheta}) includes local maxima and local minima (or ``ridges" and ``valleys") throughout the field space. See Fig.~\ref{fig:VE}. As we describe in Section \ref{sec:trajectories}, this structure of the potential yields strong single-field attractor behavior \cite{Kaiser:2012ak,Kaiser:2013sna,Schutz:2013fua,Kaiser:2015usz,DeCross:2015uza}: the system generically settles into a local minimum of the potential very quickly after the start of inflation and remains within that minimum for the duration of inflation.

Potentials of the form in Eq.~(\ref{VErtheta}) have very flat plateaus at large field values, of the type favored by recent measurements of CMB anisotropies \cite{Planck:2018jri}. For models in which $\xi_\phi \simeq \xi_\chi$, in the limit in which the ${\cal D} (\theta) r^4$ term dominates the numerator of $V (r, \theta)$ and $\xi_\phi r^2 \gg M_{\rm pl}^2$, the potential reduces to the simple form
%%%%%
\beq
V (r, \theta) \simeq \frac{ M_{\rm pl}^4 \, {\cal D} (\theta) }{\xi_\phi^2} + {\cal O} \left( \frac{ M_{\rm pl}^2}{\xi_\phi r^2} \right) .
\label{Vplateau}
\eeq
In the absence of strong turning among the background fields during inflation ($\omega^2 \ll H^2$), the upper bound on the primordial tensor-to-scalar ratio $r_{0.05} < 0.036$ at the CMB pivot scale $k_* = 0.05 \, {\rm Mpc}^{-1}$ \cite{BICEP:2021xfz} constrains $H (t_*) < 1.9 \times 10^{-5} \, M_{\rm pl}$. This constraint on $H (t_*)$ becomes more complicated for inflationary trajectories that feature strong turning before the end of inflation \cite{McDonough:2020gmn}, but is appropriate for the scenarios we consider here. Assuming that the CMB-relevant curvature perturbations crossed outside the Hubble radius while the fields were still on the large-field plateau of the potential, the constraint on $H(t_*)$ corresponds to the limit
%%%%%%
\beq
\frac{ {\cal D} (\theta) }{\xi_\phi^2} \leq 1.1 \times 10^{-9} ,
\label{DxiCMB}
\eeq
upon relating $H$ to $V$ during slow roll. From Eq.~(\ref{BCDdef}) we see that ${\cal D} (\theta) \sim 9 c_{\rm max}^2$, where $c_{\rm max} = {\rm max} \{ c_i \}$. Hence to remain compatible with observations of the CMB, we expect the couplings to fall within a range such that
%%%%%
\beq
\frac{ \vert c_{\rm max} \vert}{\xi_\phi} \lesssim {\cal O} (10^{-5}) .
\label{cmax}
\eeq
As $\xi_\phi \simeq \xi_\chi$ becomes larger, the dimensionless couplings $c_i$ can likewise become larger while still remaining compatible with observations.

The Einstein-frame potential $V (r, \theta)$ of Eq.~(\ref{VErtheta}) retains the large-field plateau as in the models studied in Refs.~\cite{Kaiser:2012ak,Kaiser:2013sna,Schutz:2013fua,Kaiser:2015usz}. On the other hand, the potential of Eq.~(\ref{VErtheta}) includes modified {\it small-field} structure compared to the previous models. In particular, the coefficients ${\cal B} (\theta)$ and ${\cal C} (\theta)$ remain nonzero when at least one of the dimensionless couplings $b_i \neq 0$. These changes to the small-field structure of the potential can yield a phase of ultra-slow-roll evolution near the end of inflation, which in turn can produce PBHs.

\subsection{Inflationary Trajectories}
\label{sec:trajectories}

\begin{figure}[h]
    \centering
    \includegraphics[width=0.43\textwidth]{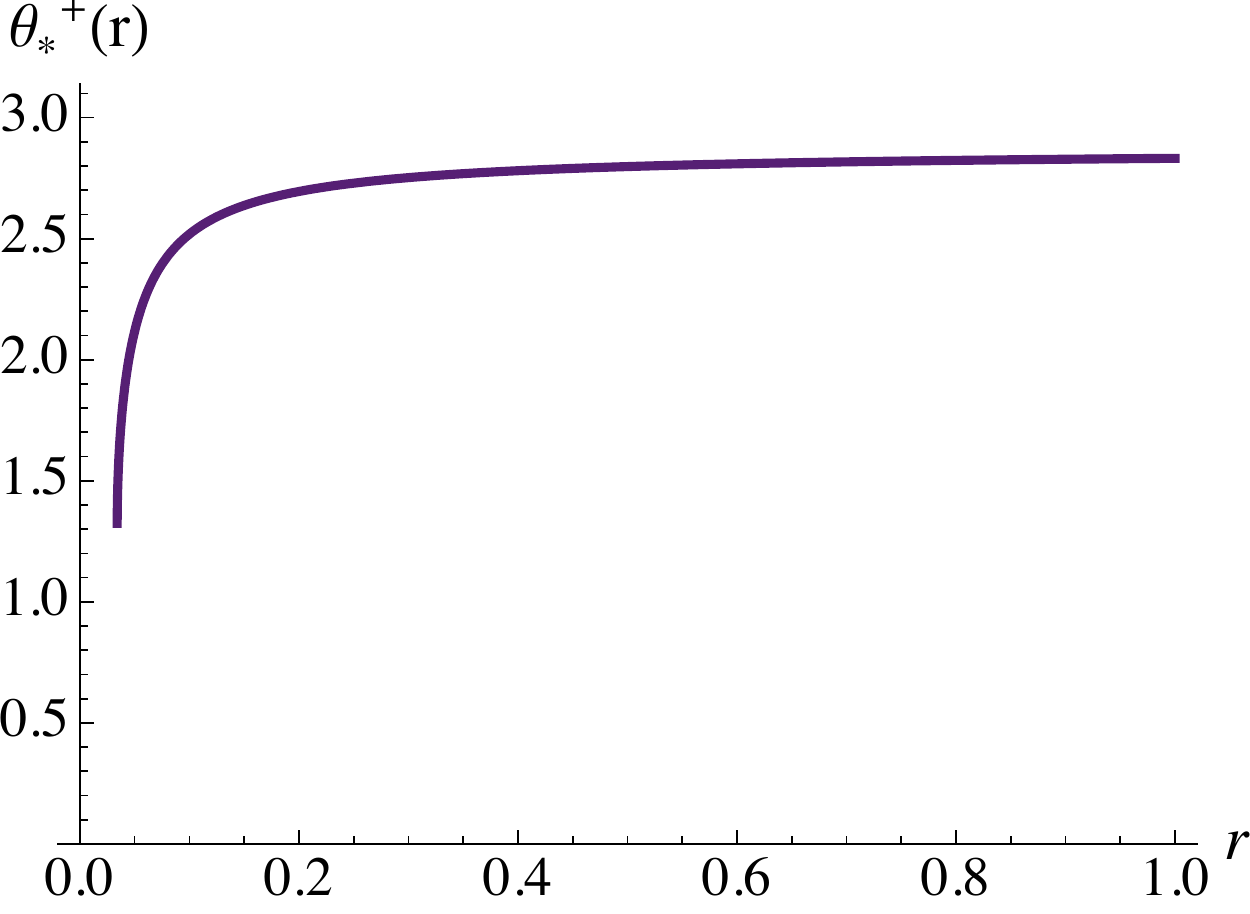}
    \caption{The angle in field space $\theta_* (r)$ along which the system evolves for the same couplings as in Fig.~\ref{fig:VE}. For this set of parameters, the local minimum of the potential lies along $\theta_*^+ (r)$, whereas $\theta_*^- (r)$ is a local maximum.}
    \label{fig:thetastar}
\end{figure}

If the dimensionless couplings that appear in Eqs.~(\ref{BCDdef})--(\ref{VErtheta}) obey additional symmetries, namely
\beq
\xi_\phi = \xi_\chi = \xi, \>\> b_1 = b_2 = b, \>\>  c_2 = c_3 ,
\label{bcxisymmetries}
\eeq
then we may find exact analytic solutions for the background fields' trajectory during inflation. In particular, if the couplings obey the relationships of Eq.~(\ref{bcxisymmetries}), then we find
%%%%%
\beq
V_{, \theta} (r, \theta) = \frac{ M_{\rm pl}^4 r^3}{[2 f (r) ]^2} \left[ {\cal C}' (\theta) \mu  + {\cal D}' (\theta) r \right]
\label{Vprime}
\eeq
because $f (r, \theta) \rightarrow f (r)$ and ${\cal B} (\theta) \rightarrow 4 b^2$ when $\xi_\phi = \xi_\chi$ and $b_1 = b_2 = b$. The system will evolve along a direction in field space $\theta_*$ such that $V_{, \theta} (r, \theta_*) = 0$. As shown in Appendix \ref{appTheta}, for the symmetric couplings of Eq.~(\ref{bcxisymmetries}) the extrema are given by
%%%%%
\beq
\theta_*^\pm (r) = {\rm arccos} (x^\pm (r))
\label{thetastar1}
\eeq
with
%%%%%
\beq
x^\pm (r) = \frac{ -d_1 \pm \vert d_4 \vert \sqrt{ - 1 + R^2} }{R \sqrt{ d_1^2 + d_4^2}} ,
\label{xpm}
\eeq
where
%%%%%
\beq
\begin{split}
    d_1 &\equiv c_1 + \frac{ c_2}{3} , \>\> d_4 \equiv c_4 + \frac{ c_2}{3} , \\
    r_{\rm imag} &\equiv \frac{ b\mu}{\sqrt{d_1^2+ d_4^2}} , \>\> R \equiv \frac{ r}{r_{\rm imag}} .
\end{split}
\label{didef}
\eeq
In the limit $b \rightarrow 0$, $x^\pm (r) \rightarrow {\rm constant}$ and hence $\dot{\theta}^\pm_* \rightarrow 0$, consistent with the non-turning attractor trajectories identified in Refs.~\cite{Kaiser:2012ak,Kaiser:2013sna,Schutz:2013fua,Kaiser:2015usz}. For $b \neq 0$, the trajectories $\theta_*^\pm (r)$ show virtually no turning until $r \ll M_{\rm pl}$, near the end of inflation. See Fig.~\ref{fig:thetastar}. The analytic solutions $\theta_*^\pm (r)$ become complex for $r < \vert r_{\rm imag} \vert$, although the fields' dynamical evolution remains smooth in the vicinity of $r \sim \vert r_{\rm imag} \vert$.

\begin{figure*}
    \centering
    \includegraphics[width=0.45\textwidth]{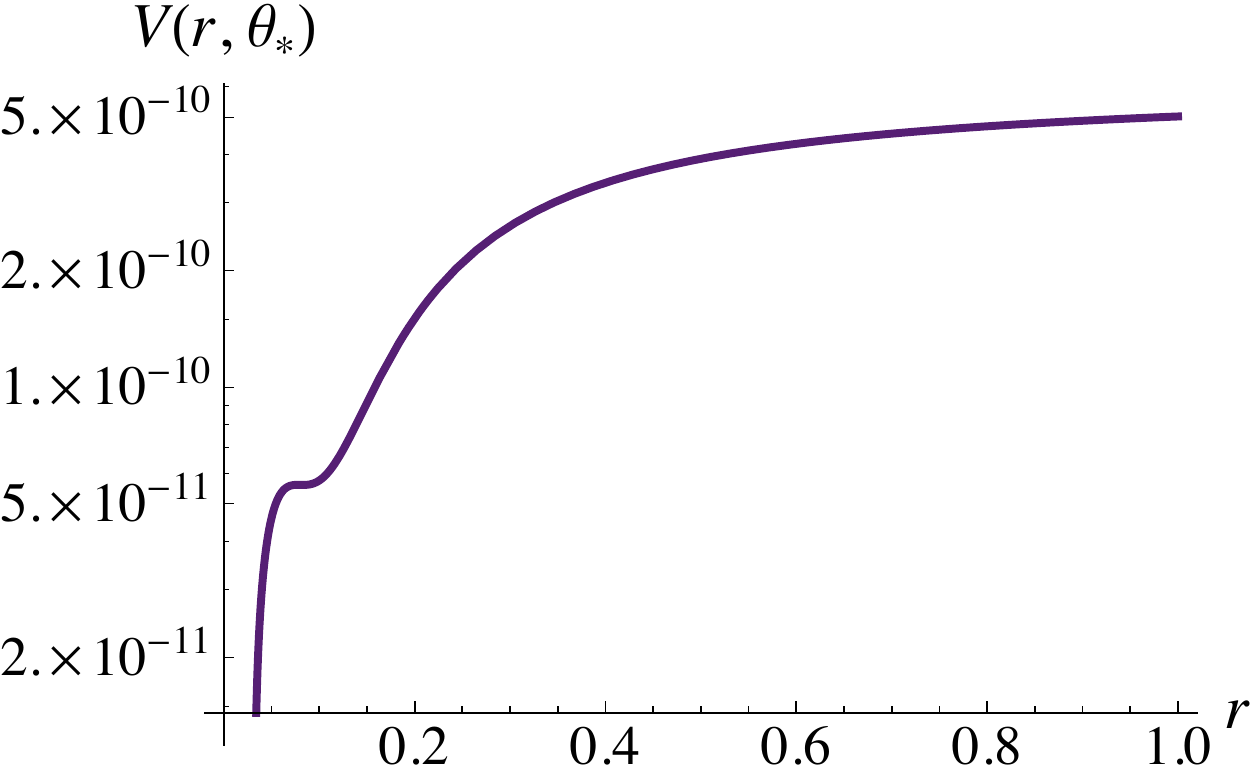} $\quad$
    \includegraphics[width=0.45\textwidth]{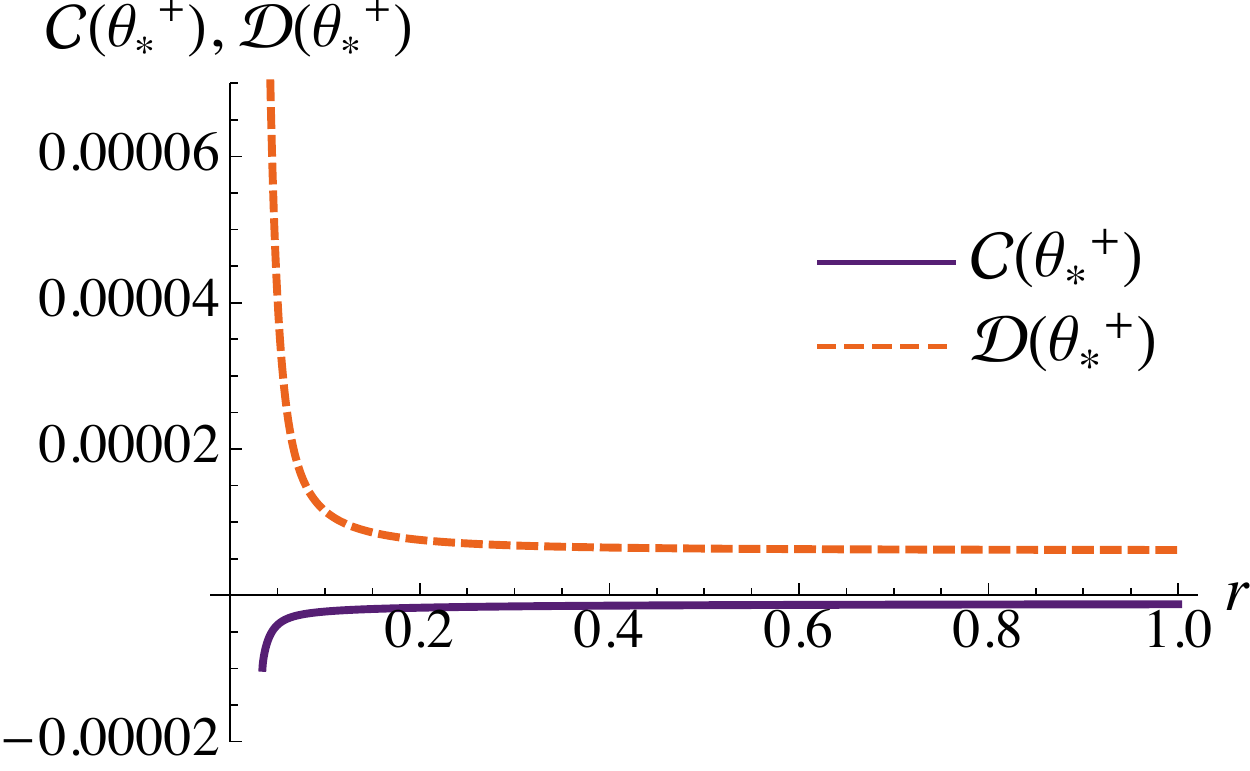}
    \caption{({\it Left}) The scalar potential in the Einstein frame $V (r, \theta_*^+)$ (in units of $M_{\rm pl}^4$) evaluated along the direction of the fields' evolution, $\theta_*^+ (r)$. ({\it Right}) The dimensionless coefficients ${\cal C} (\theta)$ (purple) and ${\cal D} (\theta)$ (orange dashed) as defined in Eq.~(\ref{BCDdef}), evaluated along the direction of the fields' evolution, $\theta_*^+ (r)$. Both plots use the same parameters as in Fig.~\ref{fig:VE}.   }
    \label{fig:VCD}
\end{figure*}

We may project the multifield potential $V (r, \theta)$ along the fields' trajectory $\theta_* (r)$, which yields $V (r, \theta_* (r))$. See Fig.~\ref{fig:VCD}. Upon including $b \neq 0$, and hence ${\cal C} \neq 0$, the potential evaluated along $\theta_* (r)$ generically develops a feature at small field values, much as in the single-field models studied in Refs.~\cite{Garcia-Bellido:2017mdw,Ezquiaga:2017fvi,Germani:2017bcs,Kannike:2017bxn}. For the example shown, the dimensionless coefficient ${\cal C} (\theta_*) < 0$ for the duration of inflation, while ${\cal B}, {\cal D} (\theta_*) > 0$ (recall that for $b_1 = b_2 = b$, ${\cal B} = 4 b^2$ is independent of $\theta$). Given the opposite signs of ${\cal C}$ and ${\cal B, D}$, the new features will emerge in $V (r, \theta_* (r))$ for field values $r$ such that $\vert {\cal C} (\theta_*)\vert \mu r \sim {\cal B} \mu^2 + {\cal D} (\theta_*) r^2$. For the parameters shown in Figs.~\ref{fig:VE}--\ref{fig:VCD}, this occurs for $r \simeq 0.1 \, \mu$.

\begin{figure*}
    \centering
    \includegraphics[width=0.48\textwidth]{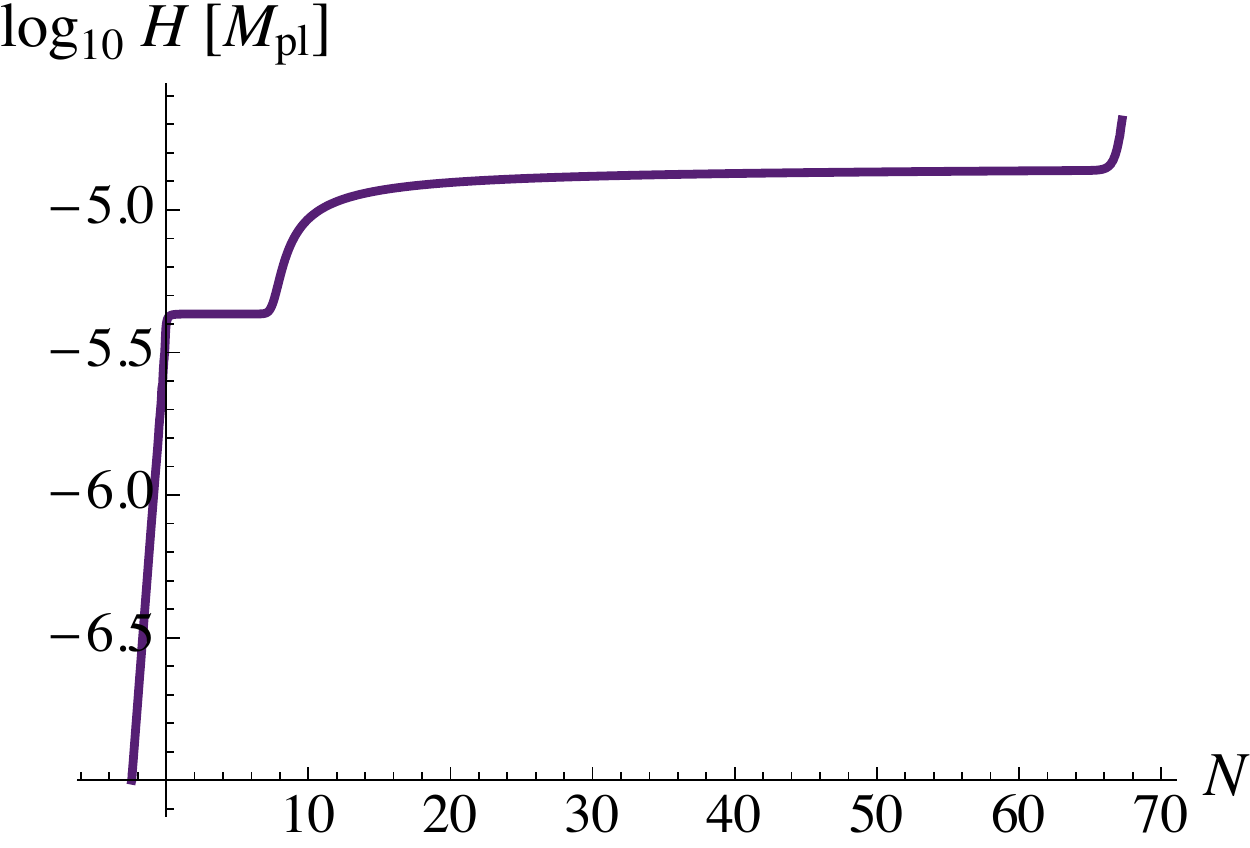} $\quad$
    \includegraphics[width=0.45\textwidth]{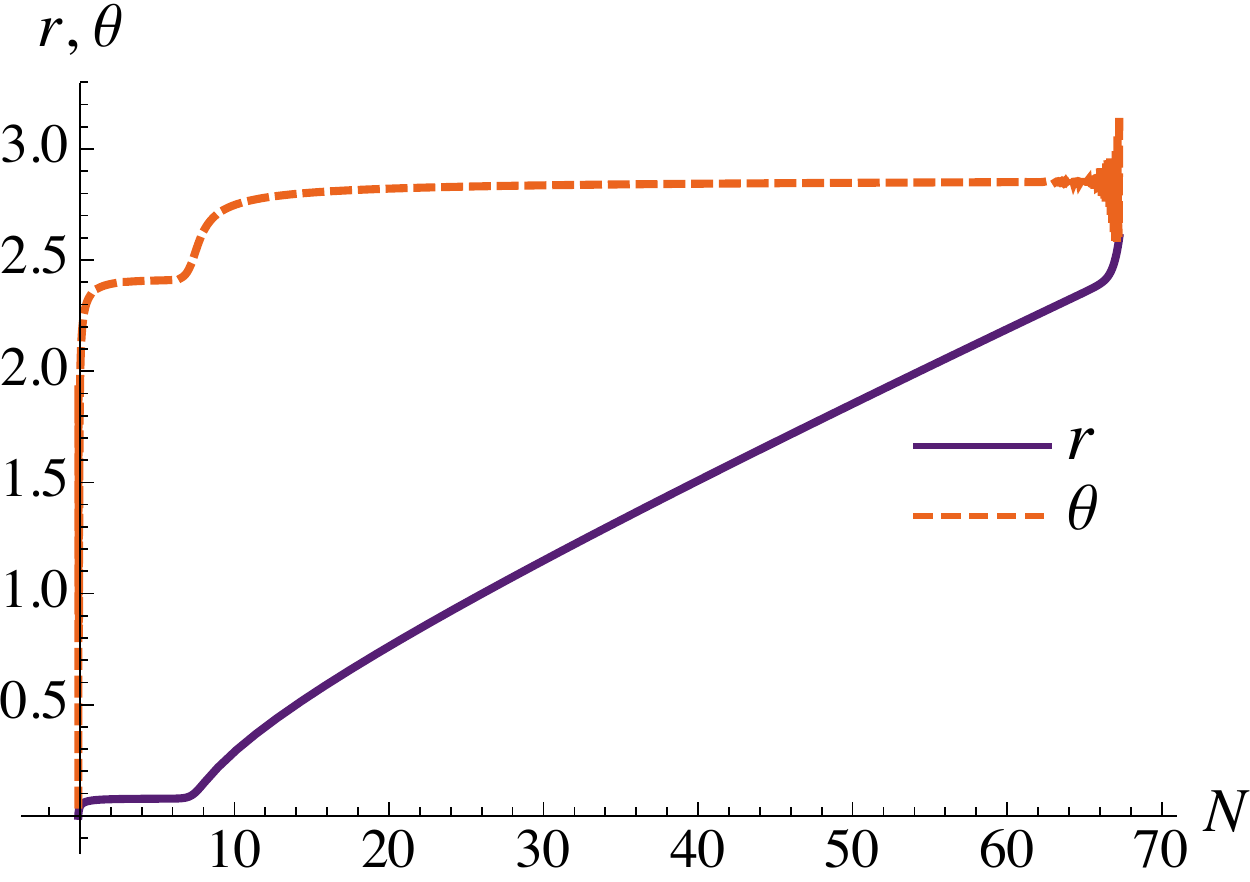}
    \caption{({\it Left}) The evolution of the Hubble parameter $H (t)$ as a function of efolds $N$ before the end of inflation ($N (t_{\rm end}) = 0$). ({\it Right}) The evolution of the fields $r (t)$ (purple, in units of $M_{\rm pl}$) and $\theta (t)$ (orange dashed) as a function of efolds $N$ before the end of inflation. Both plots use the same parameters as in Fig.~\ref{fig:VE} and initial conditions $r (t_i) = 2.6 \, M_{\rm pl}$, $\theta (t_i) = \pi - 0.02$, $\dot{r} (t_i) = - 10^{-5} \, M_{\rm pl}^2$, and $\dot{\theta} (t_i) = 4 \times 10^{-5} \, M_{\rm pl}$.  }
    \label{fig:Hfields}
\end{figure*}

\begin{figure*}
    \centering
    \includegraphics[width=0.48\textwidth]{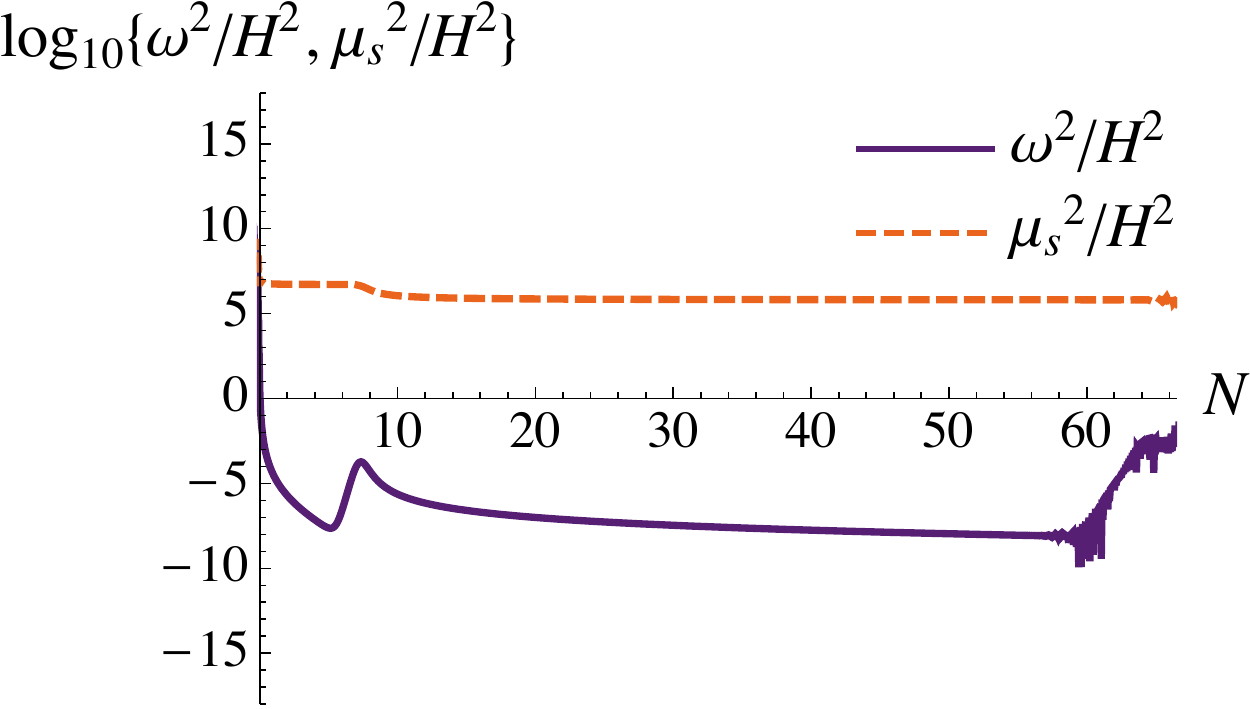} $\quad$ 
    \includegraphics[width=0.45\textwidth]{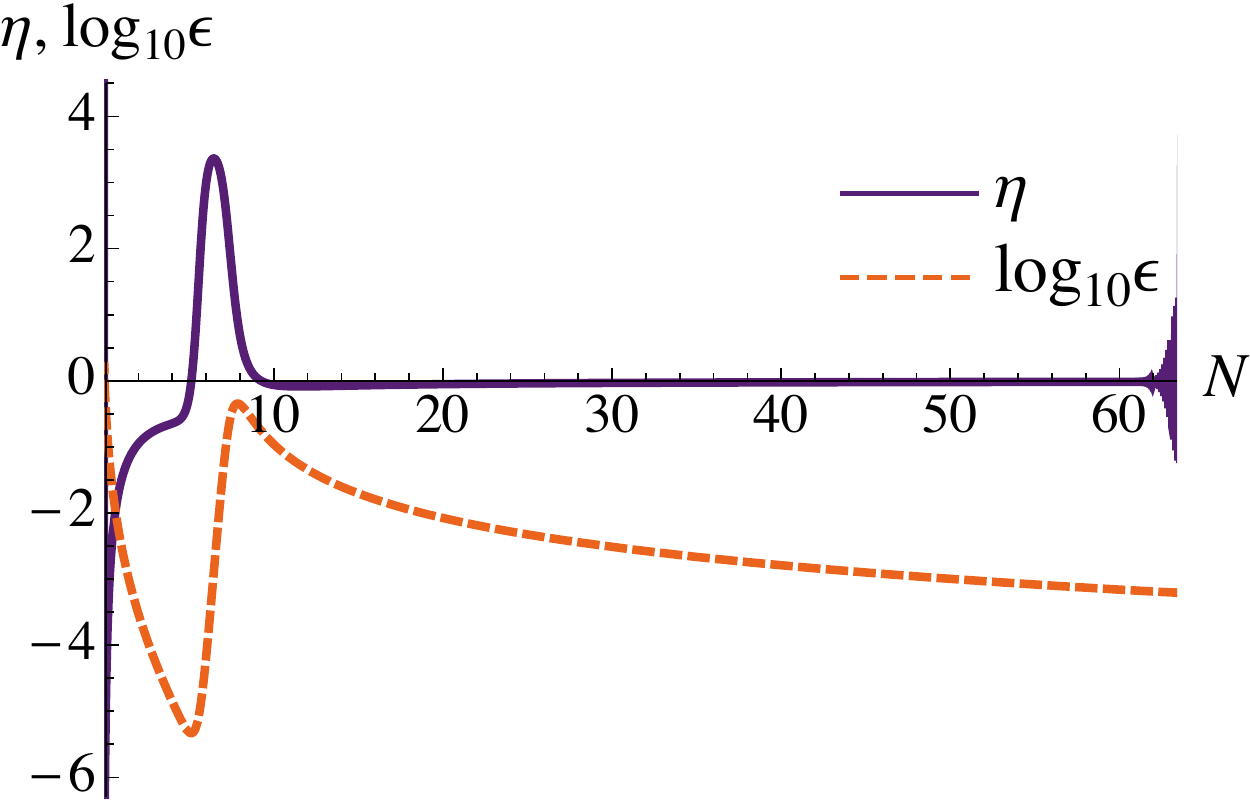}
    \caption{({\it Left}) The evolution of the covariant turn-rate $\vert \omega^I (t) \vert$ (purple) and the mass of the isocurvature modes $\mu_s (t)$ (orange dashed) as a function of efolds $N$ before the end of inflation ($N (t_{\rm end}) = 0$).  ({\it Right}) The slow-roll parameters $\eta$ (purple) and $\epsilon$ (orange dashed) as functions of efolds $N$ before the end of inflation. While the system undergoes ultra-slow-roll evolution, $\eta \rightarrow 3$ and $\epsilon \rightarrow 10^{-5}$, consistent with Eq.~(\ref{etaUSR}). Both plots use the same parameters and initial conditions as in Fig.~\ref{fig:Hfields}. 
}
    \label{fig:muomegaPR}
\end{figure*}

With fine-tuning of at least one of the couplings $\{ b_, c_i \}$, one may arrange for the small-field feature to be a quasi-inflection point, as in Refs.~\cite{Garcia-Bellido:2017mdw,Ballesteros:2017fsr,Di:2017ndc,Motohashi:2017kbs}. More generally, the projected potential will develop a local minimum along the direction $\theta_* (r)$ with a nearby local maximum, as in Ref.~\cite{Kannike:2017bxn}. When the fields encounter this small-field feature in the potential, the system enters a phase of ultra-slow-roll evolution: the fields' kinetic energy density $\rho_{\rm kin} = \dot{\sigma}^2 / 2 \rightarrow 0$ while $H \simeq {\rm constant}$, and hence $\epsilon$ falls by several orders of magnitude, given the relationship in Eq.~(\ref{epsilon}). 

We numerically solve the coupled equations of motion for the background fields $r (t), \theta (t)$ and the Hubble parameter $H (t)$ using Eqs.~(\ref{eomvarphi}) and (\ref{eombackground}). In Fig.~\ref{fig:Hfields} we plot the evolution of $H, r$ and $\theta $ for typical values of the couplings. Fig.~\ref{fig:muomegaPR} confirms that once the system settles into a local minimum of the potential in the angular direction ($V_{, \theta} (r, \theta_* (r)) = 0$), the isocurvature modes remain heavy for the duration of inflation ($\mu_s^2 \gg H^2$) and the turn-rate remains negligible ($\omega^2 \ll H^2$). When the fields encounter the small-field feature in the potential near $r \simeq 0.1 \, \mu$, the system enters a phase of ultra-slow-roll evolution, with $\eta \rightarrow 3$ and $\epsilon \rightarrow 10^{-5}$. For each of these plots, we show the evolution of the system as a function of the number of efolds $N$ before the end of inflation: $N (t) \equiv N_{\rm total} - \int_{t_i}^{t} H (t) \, dt$, where $N_{\rm total} \equiv \int_{t_i}^{t_{\rm end}} H (t) \, dt$ and $t_{\rm end}$ is determined via $\epsilon (t_{\rm end}) = 1$.

\begin{figure}[t]
    \centering
    \includegraphics[width=0.45\textwidth]{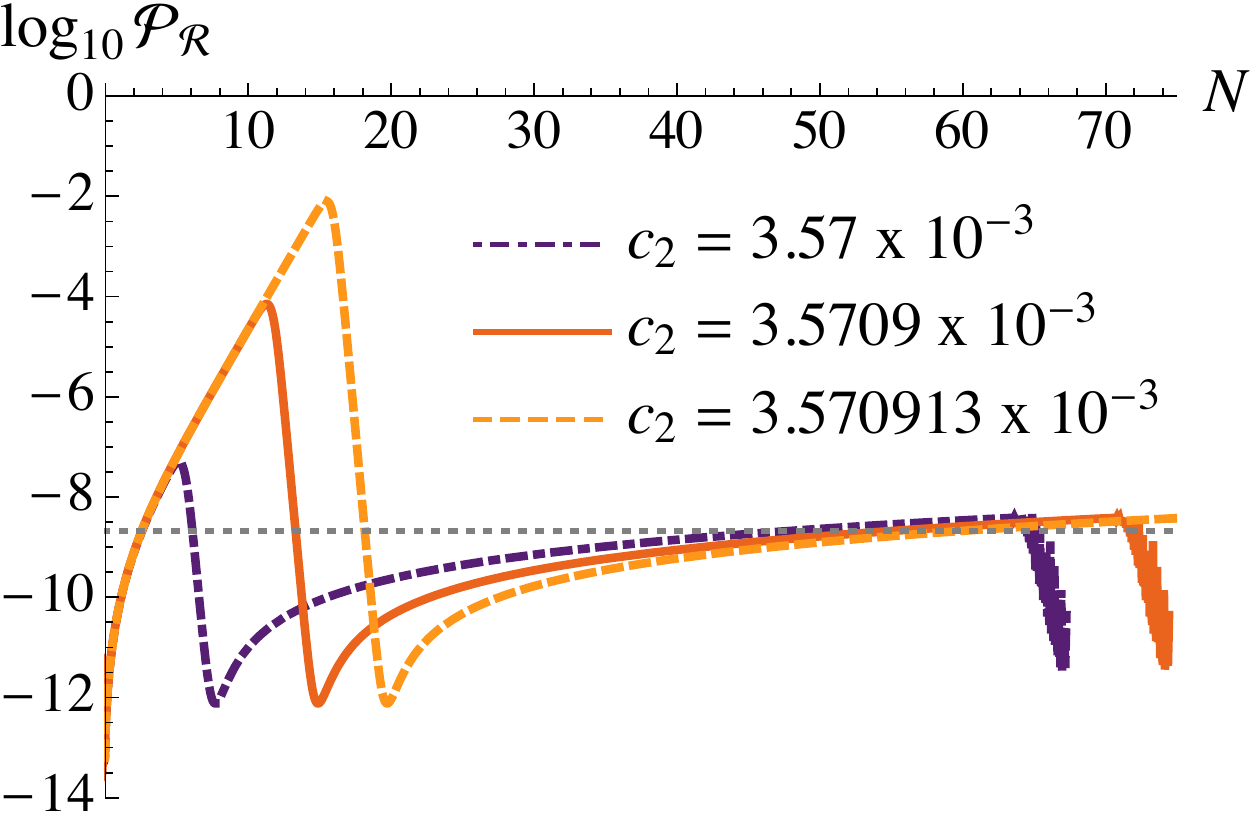}
    \caption{Fine-tuning one of the dimensionless couplings can increase the duration of the ultra-slow-roll phase. For longer periods of ultra-slow-roll, the slow-roll parameter $\epsilon$ falls to smaller values and the peak in the power spectrum ${\cal P}_{\cal R} (k)$ rises. All three curves shown here use the same parameters and initial conditions as in Figs.~\ref{fig:VE}--\ref{fig:muomegaPR}, with increasing fine-tuning of $c_2 = c_3$. The horizontal dotted line shows the COBE normalization ${\cal P}_{\cal R} (k_*) = 2.1 \times 10^{-9}$ for the CMB pivot-scale $k_* = 0.05 \, {\rm Mpc}^{-1}$.}
    \label{fig:PRtuning}
\end{figure}

Given the relationship between ${\cal P}_{\cal R} (k)$, $H$, and $\epsilon$ in Eq.~(\ref{PRHepsilon}), the power spectrum of curvature perturbations will become amplified for modes $k$ that exit the Hubble radius while the fields are in the phase of ultra-slow-roll. In general, the decrease in $\epsilon$---and hence the increase in ${\cal P}_{\cal R} (k)$---depends on the ratios of various couplings. For the parameters shown in Figs.~\ref{fig:VCD}--\ref{fig:muomegaPR}, the local maximum of the potential near $r \simeq 0.1 \, \mu$ is marginally greater than the value of the potential at the nearby local minimum, so the system spends only $\Delta N \sim 2.5$ efolds in the ultra-slow-roll phase. As shown in Fig.~\ref{fig:PRtuning}, by fine-tuning one of the dimensionless couplings, we may adjust the relative heights of the local maximum and local minimum along $\theta_* (r)$, thereby prolonging the duration over which the fields persist in the ultra-slow-roll phase and increasing the peak value of ${\cal P}_{\cal R} (k)$. Even the tallest peak of ${\cal P}_{\cal R} (k)$ shown in Fig.~\ref{fig:PRtuning} satisfies ${\cal P}_{\cal R} (k) \lesssim 10^{-2} < 1/6$, and hence the criterion of Eq.~(\ref{epsUSRdominate}) is always satisfied. In other words, even while the system undergoes ultra-slow-roll evolution, the classical evolution of the background fields dominates quantum diffusion for the  parameters considered here. 

The dynamics of the fields in the models we consider here are distinct from those recently studied in $\alpha$-attractor models \cite{Iacconi:2021ltm,Kallosh:2022vha}. In particular, we only consider positive values of the nonminimal couplings in this paper, so that the conformal transformation associated with the factor $\Omega^2 (x)$ in Eq.~(\ref{Omega}) remains nonsingular. For $\xi_I > 0$, the induced field-space manifold in the Einstein frame has positive curvature, ${\cal R}_{\rm fs} > 0$, the magnitude of which falls in the limit $\xi_\phi r^2 \gg M_{\rm pl}^2$. (An explicit expression for ${\cal R}_{\rm fs}$ for these models may be found in Eq.~(115) of Ref.~\cite{Kaiser:2012ak}.) Hence curved field-space effects make fairly modest contributions to the fields' dynamics during the early stages of inflation \cite{Kaiser:2012ak,Kaiser:2013sna,Schutz:2013fua,DeCross:2015uza}. 

In $\alpha$-attractor models, on the other hand, the curvature of the field-space manifold is negative and constant, ${\cal R}_{\rm fs} = - 4 / (3 \alpha)$, with dimensionless constant $\alpha >0$. For $\alpha \sim {\cal O} (1)$, the fields' evolution will be affected by the nontrivial field-space manifold throughout the duration of inflation. Hence in $\alpha$-attractor models, the fields may ``ride the ridge," remaining on or near a local maximum of the potential for much of the duration of inflation \cite{Iacconi:2021ltm}, whereas in the family of models we consider here, the fields generically settle into a local minimum of the potential after a brief, initial transient. For the case of $\xi_I > 0$, the fields can only ``ride the ridge" of the potential for $N \gtrsim {\cal O} (1)$ efolds if the fields' initial conditions are exponentially fine-tuned \cite{Kaiser:2012ak,Schutz:2013fua,Kaiser:2013sna,DeCross:2015uza}. The fact that the fields generically settle into a local minimum of the potential in these models ensures that the isocurvature modes remain heavy throughout inflation and that the covariant turn-rate remains negligible.

\subsection{Scaling Relationships}
\label{sec:scaling}

As shown in Fig.~\ref{fig:PRtuning}, the evolution of perturbations is sensitive to the small-field feature in the Einstein-frame potential, which in turn depends upon ratios among the dimensionless couplings $b_i$ and $c_i$. We explore some of those relationships in this section. We first note from Eqs.~(\ref{BCDdef}) and (\ref{VErtheta}) that the mass-scale $\mu$ only appears in $V (\phi^I)$ multiplied by the $b_i$. Without loss of generality, we therefore fix $\mu = M_{\rm pl}$ and adjust the magnitude of the scalar fields' tree-level masses by changing $b_i$.

\begin{figure*}[t]
    \centering
    \includegraphics[width=0.45\textwidth]{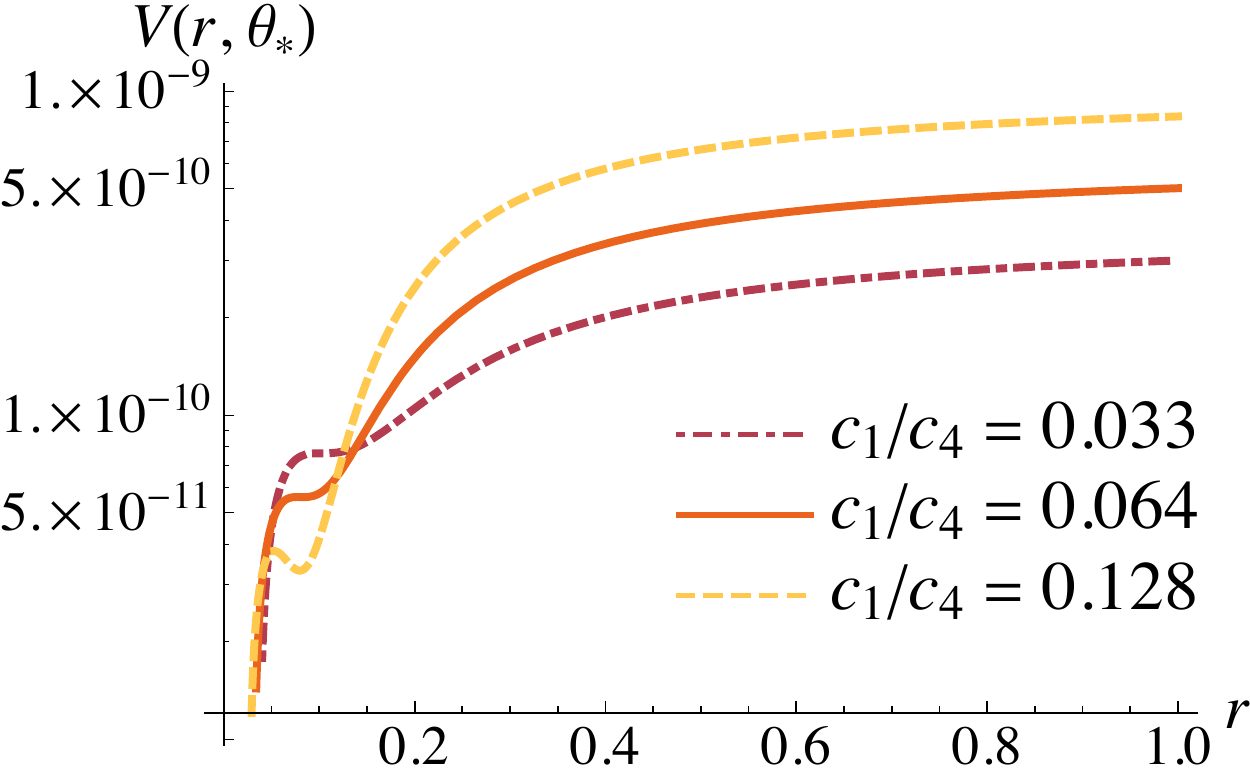} $\quad$
    \includegraphics[width=0.45\textwidth]{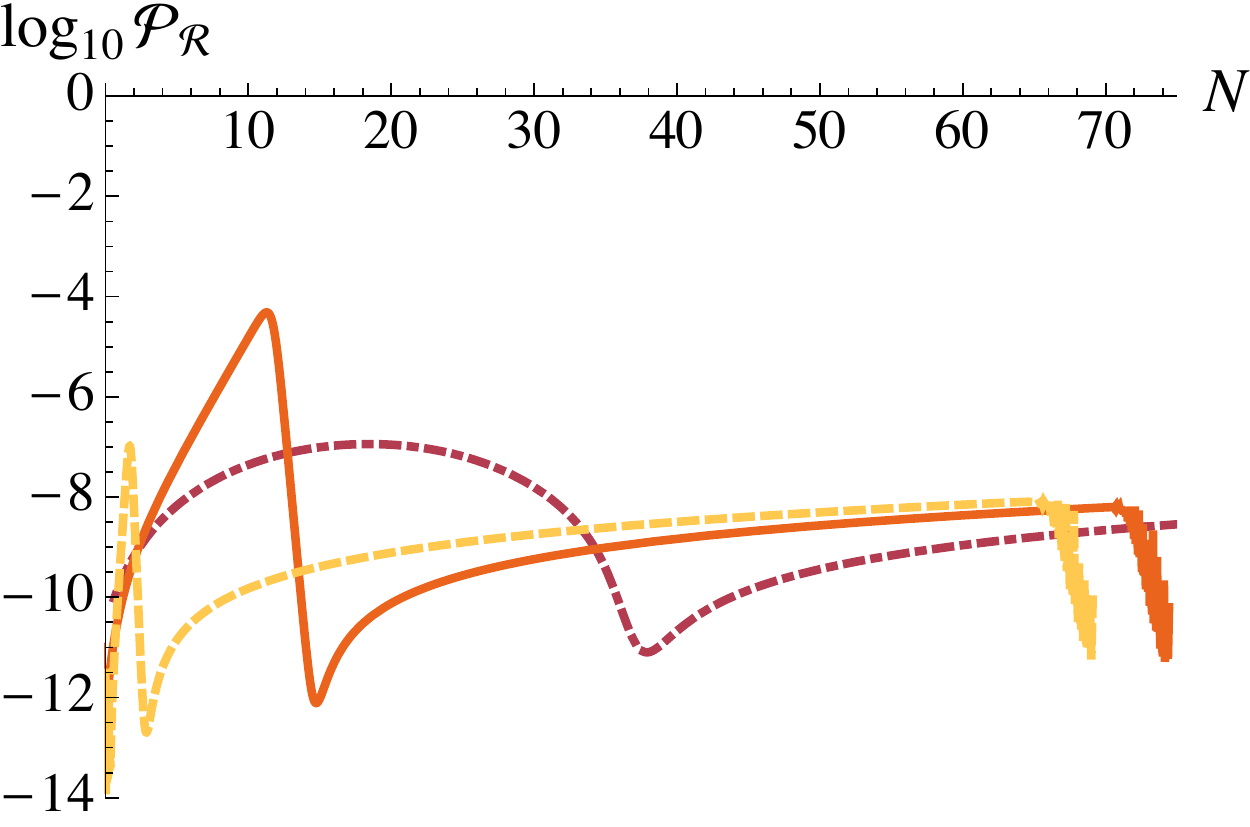}
    \caption{The potential $V (r, \theta_* (r))$ ({\it left}) and the power spectrum ${\cal P}_{\cal R} (k)$ ({\it right}) for $\mu = M_{\rm pl}$, $\xi_\phi = \xi_\chi = 100$, and $b_1 = b_2 = -1.8 \times 10^{-4}$, with varying ratio $c_1 / c_4$. In each case we keep $c_2 \sim c_4$ and fine-tune $c_2$ to a comparable degree. As the hierarchy in $V (r, \theta_*(r))$ between the large-field plateau and the small-field feature decreases, the peak in the power spectrum shifts from tall and narrow to short and wide. The curves shown here correspond to $\{ c_1, c_2, c_4 \} = \{ 1.5 \times 10^{-4}, 4.3738 \times 10^{-3}, 4.5 \times 10^{-3} \}$ (maroon dot-dashed), $\{ 2.5 \times 10^{-4}, 3.5709 \times 10^{-3}, 3.9 \times 10^{-3} \}$ (orange), and $\{ 4.1 \times 10^{-4}, 3.0879 \times 10^{-3}, 3.2 \times 10^{-3} \}$ (gold dashed).}
    \label{fig:PRcratio}
\end{figure*}

The shape of the peak in the power spectrum ${\cal P}_{\cal R}(k)$ depends on the hierarchy between the value of the potential $V (r, \theta_* (r))$ along the large-field plateau and in the vicinity of the small-field feature. This hierarchy, in turn, depends on the ratio of various coupling constants. For example, if the couplings satisfy the symmetries of Eq.~(\ref{bcxisymmetries}), we may hold $\xi$ and $b$ fixed and vary the ratio $c_1 / c_4$. If $c_1 \ll c_4$, then $V$ will develop a significant hierarchy between large and small field values, and the system will approach the small-field feature with correspondingly greater kinetic energy, much as analyzed in Ref.~\cite{Kannike:2017bxn} for similar single-field models. For $c_1 \ll c_4$, even if the value of $V$ at the local minimum is significantly lower than the value at the nearby local maximum, the system can nonetheless ``escape" to the global minimum of $V$ without lingering arbitrarily long near the small-field feature of the potential. In these scenarios, the corresponding peak in ${\cal P}_{\cal R} (k)$ is tall and narrow. In this paper we 
set aside the question of whether the fields could tunnel through the local barrier more quickly than they would simply flow beyond the local maximum classically.

As the ratio $c_1 / c_4$ becomes less extreme, the small-field feature in the potential more closely resembles a quasi-inflection point, akin to those studied in Ref.~\cite{Garcia-Bellido:2017mdw}. In this case, the fields approach the small-field feature with less kinetic energy and linger longer in the ultra-slow-roll phase. The resulting feature in ${\cal P}_{\cal R} (k)$ is more rounded and wide. See Fig.~\ref{fig:PRcratio}.

\begin{figure}[h!]
    \centering
    \includegraphics[width=0.45\textwidth]{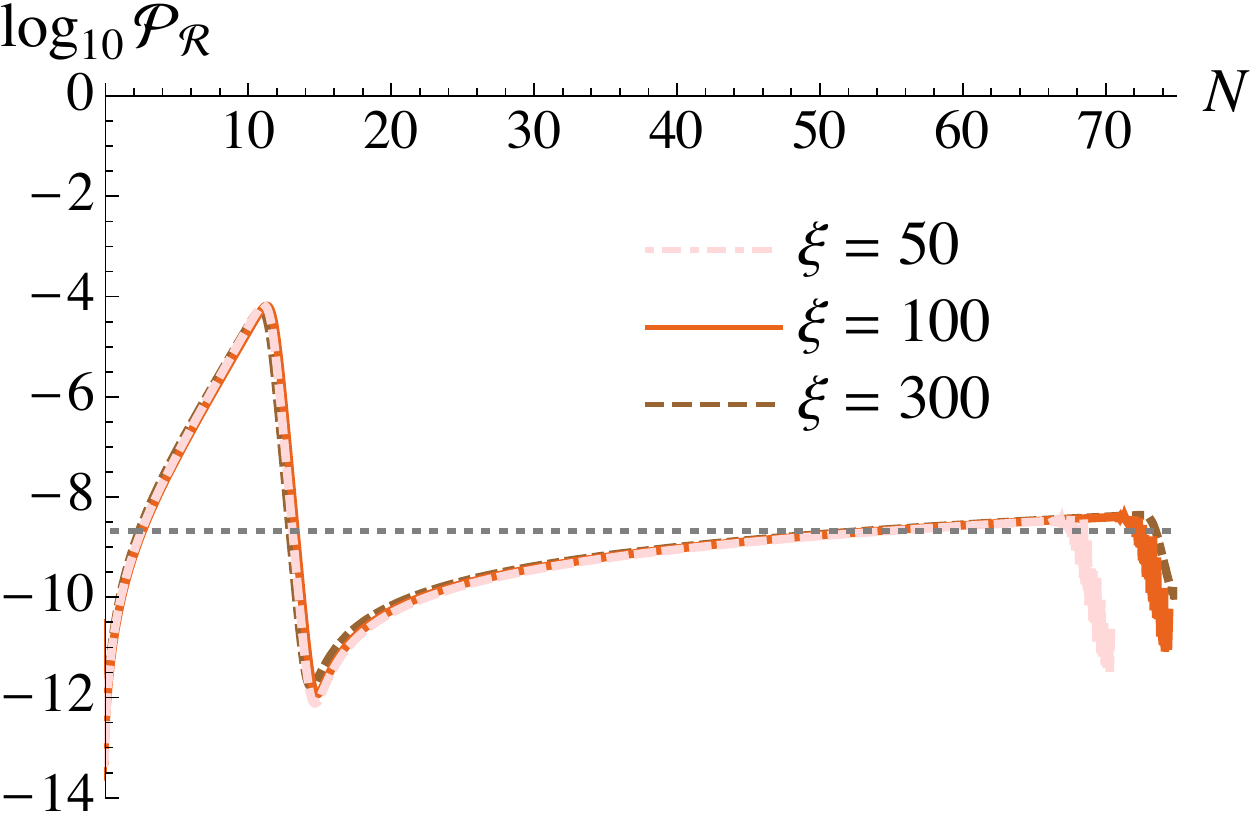}
    \caption{The power spectrum ${\cal P}_{\cal R} (k)$ for three values of the nonminimal coupling constant $\xi_\phi = \xi_\chi = \xi$, when we exploit the scaling relationships of Eqs.~(\ref{bcscaling1})--(\ref{bcscaling2}). For each curve we set $\mu = M_{\rm pl}$, $\hat{c}_1 = 2.5 \times 10^{-4}$, $\hat{c}_2 = 3.5709 \times 10^{-3}$, and $\hat{c}_4 = 3.9 \times 10^{-3}$. For $\xi = 100$ (orange) we set $y = 1$ and $\hat{b} = - 1.8 \times 10^{-4}$, and then appropriate values of $y$ and $\hat{b}$ for $\xi = 50$ (pink dot-dashed) and $\xi = 300$ (brown dashed) follow from Eq.~(\ref{bcscaling2}).}
    \label{fig:PRscaling}
\end{figure}

When the couplings obey the symmetries of Eq.~(\ref{bcxisymmetries}), the Einstein-frame potential displays a formal scaling property in the limit $\xi \gg 1$. In particular, we may set
%%%%%%
\beq
b = y \hat{b} , \>\> c_i = y \hat{c}_i 
\label{bcscaling1}
\eeq
where $y > 0$ is some constant. Note that the nonminimal coupling $\xi$ is not rescaled by $y$. Then if we fix
%%%%%
\beq
\hat{b} \sqrt{\xi} = {\rm constant} , \>\> \frac{ \xi}{y} =  {\rm constant} ,
\label{bcscaling2}
\eeq
the potential $V (r, \theta_* (r))$ is unchanged when plotted as a function of $\hat{r} \equiv r / \sqrt{\xi}$. This self-similarity, in turn, yields identical power spectra. See Fig.~\ref{fig:PRscaling}.

 Our model does not require the symmetries among coupling constants identified in Eq.~(\ref{bcxisymmetries}); in general one may consider $\xi_\phi \neq \xi_\chi$, $b_1 \neq b_2$, and/or $c_2 \neq c_3$. Relaxing the symmetries of Eq.~(\ref{bcxisymmetries}) affects the shape of the potential, especially in the vicinity of the small-field feature, which in turn can affect the fields' dynamics. We defer an exploration of this expanded parameter space to future work.

\section{PBH Formation}
\label{sec:PBHform}

PBHs can form soon after the end of inflation from large peaks in the power spectrum ${\cal P}_{\cal R} (k)$ on length-scales much shorter than those probed by the CMB. Such large perturbations cross outside the Hubble radius near the end of inflation, remain effectively frozen in amplitude while their wavelength is longer than the Hubble radius, and later re-enter the Hubble radius after the end of inflation, whereupon they can induce gravitational collapse. 

\subsection{Critical Collapse}

Upon re-entering the Hubble radius after inflation, local overdensities
%%%%%%
\beq
\delta ({\bf x}) \equiv \frac{ \rho ({\bf x} ) - \bar{\rho} }{\bar{\rho}}
\label{deltadef}
\eeq
will induce gravitational collapse if they are of sufficient amplitude. Here $\bar{\rho} = \rho_{\rm total}$ is the energy density averaged over a Hubble volume. The collapse process is a critical phenomenon akin to other kinds of phase transitions. In particular, the masses of black holes that form at time $t_c$ follow the distribution \cite{Choptuik:1992jv,Evans:1994pj,Gundlach:2002sx,Niemeyer:1997mt,Niemeyer:1999ak,Yokoyama:1998xd,Green:1999xm,Green:2004wb,Kuhnel:2015vtw,Young:2019yug,Kehagias:2019eil,Escriva:2019phb,DeLuca:2020ioi,Musco:2020jjb,Escriva:2021aeh}
%%%%
\beq
M (\delta_{\rm avg}) = {\cal K} M_H (t_c) \left( \delta_{\rm avg} - \delta_c \right)^\nu \label{Micritical}
\eeq
for overdensities $\delta_{\rm avg}$ above some threshold $\delta_c \sim {\cal O} (10^{-1})$, where $\delta_{\rm avg}$ is the spatial average of $\delta ({\bf x})$ over a region of radius $R < H^{-1}$, ${\cal K}$ is a dimensionless ${\cal O} (1)$ constant, and $\nu$ is a universal critical exponent ($\nu \simeq 0.36$ for collapse during a radiation-dominated era). The Hubble mass $M_H (t_c)$ is the mass enclosed within a Hubble sphere at time $t_c$:
%%%%%
\beq
\begin{split}
M_H (t_c) &\equiv \frac{ 4 \pi}{3} \rho_{\rm total} (t_c) H_c^{-3} \\
&= 4 \pi \frac{ M_{\rm pl}^2}{ H_c} ,
\end{split}
\label{MH}
\eeq
where $H_c \equiv H (t_c)$. The second line of Eq.~(\ref{MH}) follows upon using the Friedmann equation, $H^2 = \rho_{\rm total} / (3 M_{\rm pl}^2)$. Although the relationship between the threshold $\delta_c$ and the curvature perturbation ${\cal R}$ is, in general, nonlinear and depends on the spatial profile of the overdensities \cite{Young:2019yug,Kehagias:2019eil,Escriva:2019phb,DeLuca:2020ioi,Musco:2020jjb,Escriva:2021aeh}, the threshold criterion $\delta_{\rm avg} \geq \delta_c$ for the production of PBHs is typically equivalent to the threshold \cite{Young:2019yug} 
%%%
\beq
{\cal P}_{\cal R} (k_{\rm pbh}) \geq 10^{-3}, 
\label{PRthreshold}
\eeq
where ${\cal P}_{\cal R}$ is defined in Eq.~(\ref{PRdef}). The scale $k_{\rm pbh} = a (t_c) H_c$ is the comoving wavenumber of perturbations that re-enter the Hubble radius at time $t_c$ and induce collapse. 

The mass spectrum of PBHs that form via critical collapse includes a long tail for masses $M < \bar{M}$ \cite{Kuhnel:2015vtw,Young:2019yug,DeLuca:2020ioi}, though it is sharply peaked at an average value $\bar{M}$ that is remarkably close to Bernard Carr's original estimate \cite{Carr:1975qj},
%%%%%%%
\beq
\bar{M} = \gamma M_H (t_c) ,
\label{barMi}
\eeq
with dimensionless constant $\gamma \simeq 0.2$. For PBHs that form during the radiation-dominated phase, $a(t) \sim t^{1/2}$ and hence $H (t) = 1 / (2t)$, so from Eqs.~(\ref{MH}) and (\ref{barMi}) we have
%%%%%%
\beq
\bar{M} \simeq 8.1 \times 10^{37} \, {\rm g} \left( \frac{ t_c}{1 \, {\rm s} } \right) 
\label{barMiseconds}
\eeq
upon using $\gamma = 0.2$. PBHs with average masses within the range $10^{17} \, {\rm g} \leq \bar{M} \leq 10^{22} \, {\rm g}$ could account for the entire dark-matter fraction in the observable universe today while evading various observational constraints \cite{Carr:2020xqk,Green:2020jor,Villanueva-Domingo:2021spv}; this corresponds to PBH formation times of $10^{-21} \, {\rm s} \leq t_c \leq 10^{-16} \, {\rm s}$.

We may relate the time $t_c$ to the earlier time $t_{\rm pbh}$, during inflation, when perturbations with wavenumber $k_{\rm pbh}$ first crossed outside the Hubble radius. If the first Hubble-crossing time $t_{\rm pbh}$ occurs $\Delta N$ efolds before the end of inflation, then
%%%%%
\beq
k_{\rm pbh} = a (t_{\rm pbh}) H (t_{\rm pbh} ) = a (t_{\rm end}) e^{- \Delta N} H (t_{\rm pbh}) ,
\label{kpbh1}
\eeq
where $t_{\rm end}$ denotes the end of inflation. As in Appendix \ref{appPerturbations}, we parameterize the post-inflation reheating phase as a brief period of matter-dominated expansion ($w_{\rm eff} \simeq 0$) which lasts $N_{\rm reh}$ efolds between the times $t_{\rm end}$ and $t_{\rm rd}$; beginning at time $t_{\rm rd}$, the universe expands with a radiation-dominated equation of state \cite{Amin:2014eta,Allahverdi:2020bys}. Then the scale factor $a (t_c)$ at the time that the perturbations of comoving wavenumber $k_{\rm pbh}$ re-enter the Hubble radius will be
%%%%%
\beq
a (t_c) = a (t_{\rm end}) e^{N_{\rm reh} } \left( \frac{ t_c}{t_{\rm rd } } \right)^{1/2}
\label{atc}
\eeq
and the Hubble parameter will be $H (t_c) = 1/ (2 t_c)$. Between $t_{\rm end}$ and $t_{\rm rd}$ the energy density redshits as $\rho (t_{\rm rd}) = \rho (t_{\rm end}) e^{-3 N_{\rm reh}}$, so we may write
%%%%%
\beq
\frac{1}{ t_{\rm rd} } = 2 H (t_{\rm end}) e^{-3 N_{\rm reh} / 2} .
\label{tRD}
\eeq
From Eqs.~(\ref{atc}) and (\ref{tRD}), we find
%%%%%
\beq
\begin{split}
k_{\rm pbh} &= a (t_c) H (t_c) \\
&= \frac{1}{\sqrt{2 t_c}} a (t_{\rm end}) H^{1/2} (t_{\rm end}) e^{N_{\rm reh} / 4}  .
\end{split}
\label{kpbh2}
\eeq
Equating the expressions for $k_{\rm pbh}$ in Eqs.~(\ref{kpbh1}) and (\ref{kpbh2}), we may solve for $\Delta N$:
%%%%%
\beq
\Delta N = \frac{1}{2} \ln \left[  \frac{ 2 H^2 (t_{\rm pbh} ) }{H (t_{\rm end} ) } e^{- N_{\rm reh} / 4}\, t_c \right] .
\label{DeltaN}
\eeq
For the parameters that we have been considering, which yield a substantial hierarchy between the values of the potential along the large-field plateau and near the small-field feature, $H (t_{\rm pbh} ) \simeq H (t_{\rm end} ) \simeq 10^{-5.4} \, M_{\rm pl}$; see the left panel of Fig.~\ref{fig:Hfields}. Previous studies of post-inflation reheating in closely related models have consistently found efficient reheating, with $N_{\rm reh} \lesssim 3$ across a wide range of parameter space \cite{DeCross:2015uza,DeCross:2016cbs,DeCross:2016fdz,Nguyen:2019kbm,vandeVis:2020qcp}; the incorporation of trilinear couplings, such as the terms proportional to the coefficient ${\cal C}$ in the effective potential of Eq.~(\ref{VErtheta}), generically increases the efficiency of reheating \cite{Bassett:1999ta,Dufaux:2006ee}. Upon taking $0 \leq N_{\rm reh} \leq 3$, we therefore find
\beq
18 \lesssim \Delta N \lesssim 25 
\label{DeltaNtarget}
\eeq
across the range of PBH formation times of interest, $10^{-21} \, {\rm s} \leq t_c \leq 10^{-16} \, {\rm s}$. 

\subsection{PBHs from Ultra-Slow-Roll Evolution in These Models}
\label{sec:PBHsUSR}

\begin{figure*}
    \centering
    \includegraphics[width=0.43\textwidth]{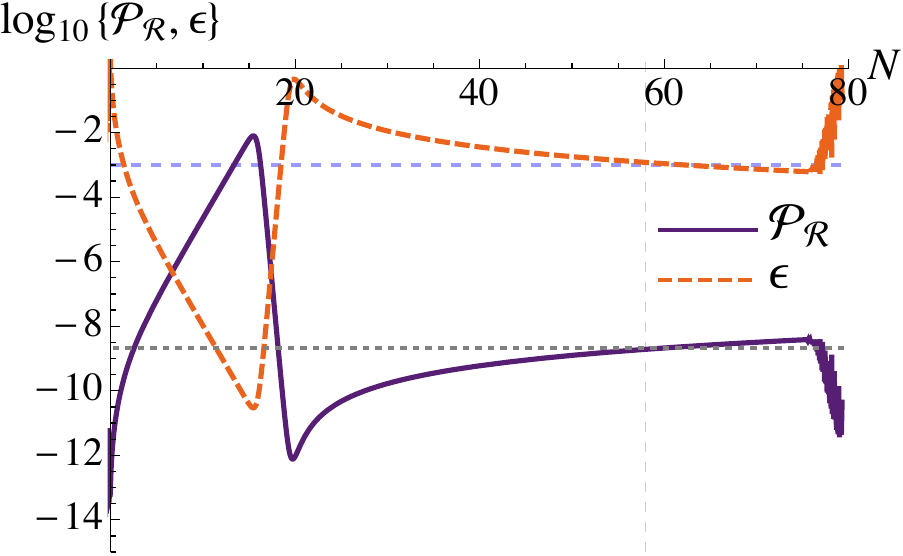} $\quad$ \includegraphics[width=0.43\textwidth]{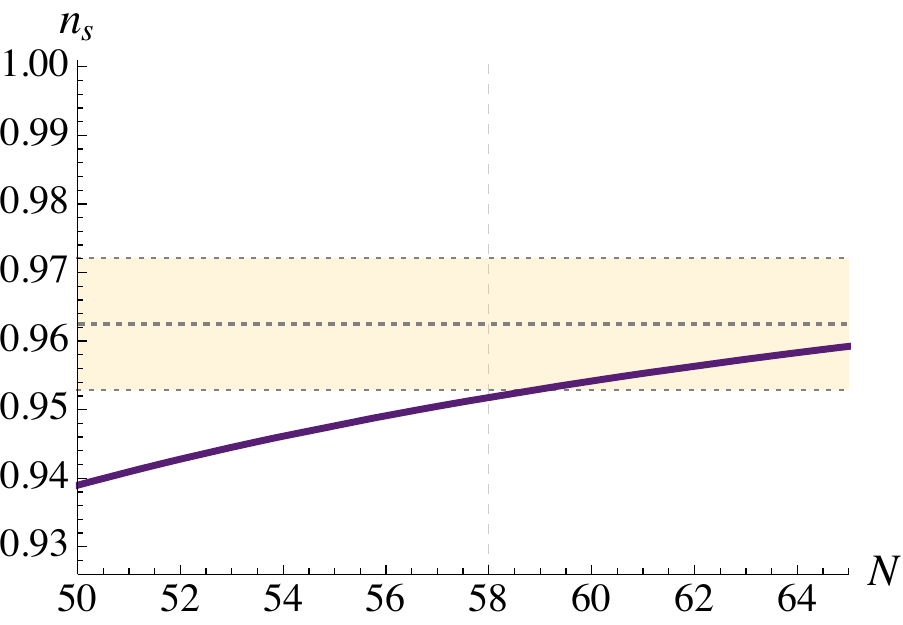} \\
    \includegraphics[width=0.43\textwidth]{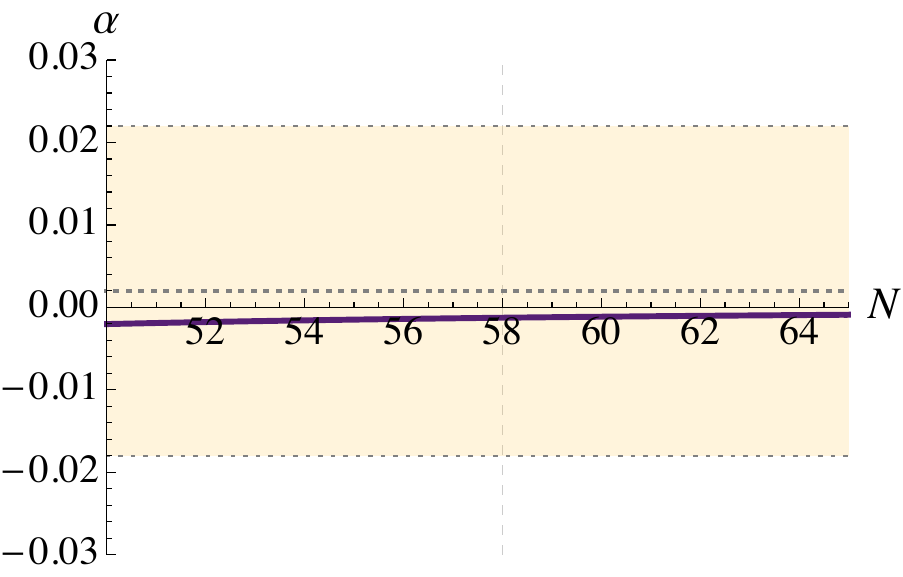} $\quad$ \includegraphics[width=0.43\textwidth]{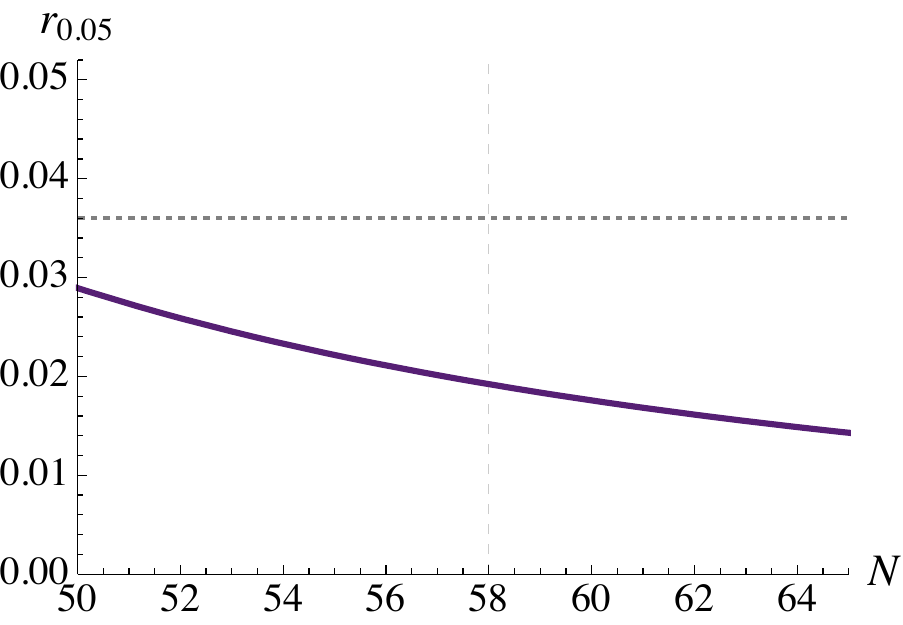}    
    \caption{Observable quantities from our two-field model with one fine-tuned parameter. For each plot, $N$ denotes the number of efolds before the end of inflation ($N (t_{\rm end}) = 0$). For the parameters chosen, the CMB pivot scale $k_* = 0.05 \, {\rm Mpc}^{-1}$ crossed outside the Hubble radius $N_* \simeq 58$ efolds before the end of inflation, and ${\cal P}_{\cal R} (k)$ first exceeded the threshold for PBH production $\Delta N = 16.3$ efolds before the end of inflation. ({\it Top left}) The power spectrum ${\cal P}_{\cal R} (k)$ (purple) and the slow-roll parameter $\epsilon$ (orange dashed). The horizontal dotted line shows the COBE normalization ${\cal P}_{\cal R} (k_*) = 2.1 \times 10^{-9}$, and the horizontal dashed blue line shows the threshold for PBH formation ${\cal P}_{\cal R} (k) = 10^{-3}$. ({\it Top right}) The spectral index $n_s (k_*)$ (purple), Planck 2018 best-fit value (dotted), and $2 \sigma$ error-bar contours \cite{Planck:2018jri}. ({\it Bottom left}). The running of the spectral index $\alpha (k_*) = (d n_s (k) / d {\rm ln} k)\vert_{k_*}$ (purple), Planck 2018 best-fit value (dotted), and $2 \sigma$ error-bar contours when the Planck analysis allows for $\alpha (k_*) \neq 0$ \cite{Planck:2018jri}. ({\it Bottom right}) The tensor-to-scalar ratio $r (k_*)$ (purple) and the 2020 Planck-BICEP/Keck upper bound (dotted) \cite{BICEP:2021xfz}. The system was evolved numerically with the same parameters and initial conditions as in Figs.~\ref{fig:VE}--\ref{fig:muomegaPR}, but with $c_2 = c_3 = 3.570913 \times 10^{-3}$ rather than $c_2 = c_3 = 3.57 \times 10^{-3}$.   }
    \label{fig:PBHrun}
\end{figure*}

As analyzed in Refs.~\cite{Byrnes:2018txb,Carrilho:2019oqg}, a rapid rise in ${\cal P}_{\cal R} (k)$ at short wavelengths $k \sim k_{\rm pbh}$, which could induce PBHs after inflation, necessarily has an impact on the long-wavelength power spectrum in the vicinity of the CMB pivot-scale $k_*$; see also Ref.~\cite{Ando:2020fjm}. Hence there is a delicate balance required to secure predictions for observables in the vicinity of the CMB pivot scale $k_*$ that remain consistent with the latest measurements \cite{Planck:2018jri,BICEP:2021xfz,Planck:2019kim} while also arranging for ${\cal P}_{\cal R} (k_{\rm pbh}) \geq 10^{-3}$. In particular, the presence of small-field features in the potential, which can yield a large peak in ${\cal P}_{\cal R} (k)$ near $k \sim k_{\rm pbh}$, tends to modestly deform the potential along the large-field plateau, relevant for ${\cal P}_{\cal R} (k_*)$. The value of the spectral index $n_s (k_*)$ is typically lower than in related models for which little or no peak appears in ${\cal P}_{\cal R} (k)$ at small scales. 

To compare with the latest observations, we must evaluate the number of efolds before the end of inflation, $N_*$, when the CMB pivot scale $k_* = 0.05 \, {\rm Mpc}^{-1}$ first crossed outside the Hubble radius. Eq.~(\ref{Nstar}) shows that $N_*$ depends weakly on the duration of reheating. Given efficient reheating in these models \cite{DeCross:2015uza,DeCross:2016cbs,DeCross:2016fdz,Nguyen:2019kbm,vandeVis:2020qcp}, we take $N_{\rm reh} \sim 0$; then Eq.~(\ref{Nstar}) yields $N_* \simeq 58$ for the parameters of interest.

The models we consider here generically induce a small but nonzero running of the spectral index, $\alpha (k_*) \equiv (d n_s (k) / d {\rm ln} k )\vert_{k_*} \sim {\cal O} (10^{-3})$. If one includes possible running $\alpha (k_*) \neq 0$ in the analysis of the latest Planck data, then the best-fit value for the spectral index is given by $n_s (k_*) = 0.9625 \pm 0.0048$, with $\alpha (k_*) = 0.002 \pm 0.010$, each at $68\%$ confidence level \cite{Planck:2018jri}. Meanwhile, the most recent combined Planck-BICEP/Keck observations constrain the tensor-to-scalar ratio at $k_* = 0.05 \, {\rm Mpc}^{-1}$ to be $r_{0.05} < 0.036$ \cite{BICEP:2021xfz}. As shown in Fig.~\ref{fig:PBHrun}, for a particular choice of parameters our two-field model yields predictions consistent with the latest observations while also producing a peak in the power spectrum that first crosses the critical threshold ${\cal P}_{\cal R} (k_{\rm pbh}) \geq 10^{-3}$ at $\Delta N = 16.3$ efolds before the end of inflation.

The timing of the peak in ${\cal P}_{\cal R} (k)$ for the set of parameters shown in Fig.~\ref{fig:PBHrun} was calculated neglecting non-Gaussian features of the probability distribution function for large-amplitude curvature perturbations, which arise from stochastic effects such as quantum diffusion and backreaction. When such effects are incorporated self-consistently, the probability distribution function typically features more power in the tails of the distribution than a simple Gaussian---meaning that large fluctuations remain rare, but much {\it less} rare than standard calculations (of the sort we incorporate here) would suggest \cite{Byrnes:2012yx,Young:2013oia,Pattison:2017mbe,Biagetti:2018pjj,Kehagias:2019eil,Ezquiaga:2019ftu,Ando:2020fjm,Tada:2021zzj,Biagetti:2021eep}. Although it remains a topic for further research, we expect that such non-Gaussian effects would likely shift $\Delta N$ by ${\cal O} (1)$ efolds, which would bring $\Delta N$ more squarely within the range of Eq.~(\ref{DeltaNtarget}) of interest for dark matter abundances.  

Even while neglecting these non-Gaussian effects, we find that the results shown in Fig.~\ref{fig:PBHrun} require a substantial fine-tuning of one of the dimensionless coupling constants: $c_2 = 3.570913 \times 10^{-3}$, rather than the more ``reasonable" value $c_2 = 3.57 \times 10^{-3}$ that was used for the plots in Figs.~\ref{fig:VE}--\ref{fig:muomegaPR}. Such substantial fine-tuning is typical among models that produce PBHs from a phase of ultra-slow-roll evolution \cite{Garcia-Bellido:2017mdw,Ezquiaga:2017fvi,Kannike:2017bxn,Germani:2017bcs,Motohashi:2017kbs,Di:2017ndc,Ballesteros:2017fsr,Pattison:2017mbe,Passaglia:2018ixg,Byrnes:2018txb,Biagetti:2018pjj,Carrilho:2019oqg,Inomata:2021tpx,Inomata:2021uqj,Pattison:2021oen}.

Although the need for fine-tuning in such models is not new, we note nevertheless that the multifield models considered here are relatively efficient. We require such models to yield accurate predictions for {\it eight} distinct quantities; our two-field model does so using {\it six} relevant free parameters. The observable quantities to match include the spatial curvature contribution to the total energy density $\Omega_K$; the spectral index $n_s (k_*)$; the running of the spectral index $\alpha (k_*)$; the tensor-to-scalar ratio $r (k_*)$; the isocurvature fraction at the end of inflation $\beta_{\rm iso} (k_*, t_{\rm end})$; the non-Gaussianity parameter $f_{\rm NL}$; the peak amplitude of the power spectrum at short scales ${\cal P}_{\cal R} (k_{\rm pbh})$; and the time $\Delta N$ when the peak in ${\cal P}_{\cal R} (k_{\rm pbh})$ first crosses the critical threshold. 

The multifield models we explore here display strong single-field attractor behavior, with negligible turning throughout the duration of inflation, $\omega^2 \ll H^2$. Such attractor behavior means that the evolution of the system---and hence predictions for observables---is sensitive to changes in {\it one} initial condition, $r (t_i)$, rather than the other $2 {\cal N} - 1$ initial conditions required in ${\cal N}$-field models. (For example, predictions for observables in the two-field case are independent of $\dot{r} (t_i)$, $\theta (t_i)$, and $\dot{\theta} (t_i)$, unless those initial conditions are exponentially fine-tuned \cite{Kaiser:2012ak,Kaiser:2013sna,Schutz:2013fua,Kaiser:2015usz,DeCross:2015uza}.) Once $r (t_i)$ is set large enough to yield sufficient inflation (with $N_{\rm total} \geq 65$ efolds), these models generically satisfy observational constraints on $\Omega_K$. Meanwhile, as emphasized above, the single-field attractor behavior generically suppresses such typical multifield phenomena as $\beta_{\rm iso} (k_*, t_{\rm end})$ and $f_{\rm NL}$, thereby easily keeping predictions consistent with observational bounds. In particular, consistent with the discussion leading to Eq.~(\ref{betaisoheavy}), we find $\beta_{\rm iso} (k_*, t_{\rm end}) < e^{-3 N_*} \sim {\cal O} (10^{-76})$ for the parameters used in Fig.~\ref{fig:PBHrun}, compared to the current Planck bound $\beta_{\rm iso} (k_*, t_{\rm end}) \leq 0.026$ \cite{Planck:2018jri}. Likewise, from the discussion leading to Eq.~(\ref{fNLSFA2}), we find $f_{\rm NL}^{\rm equil} (k_*) = - 0.019$ for the parameters used in Fig.~\ref{fig:PBHrun}, consistent with the latest measurement from Planck: $f_{\rm NL}^{\rm equil}(k_*) = -26 \pm 47$ \cite{Planck:2019kim}.

The results shown in Fig.~\ref{fig:PBHrun}, which incorporate the symmetries among coupling constants of Eq.~(\ref{bcxisymmetries}), thus reveal close agreement between predictions for $\{ \Omega_K, \beta_{\rm iso}, f_{\rm NL}, n_s (k_*), \alpha (k_*), r (k_*), {\cal P}_{\cal R} (k_{\rm pbh}), \Delta N \}$ from a two-field model with six relevant free parameters: $\{ r (t_i), \xi, b, c_1, c_2, c_4 \}.$

\section{Discussion}
\label{sec:discussion}

In this paper we have demonstrated that inflationary models that incorporate well-motivated features from high-energy physics can produce primordial black holes (PBHs) soon after the end of inflation, of interest for present-day dark-matter abundances. In particular, we have investigated models with multiple interacting scalar fields, each with a nonminimal coupling to the spacetime Ricci curvature scalar. Our multifield models are inspired by supersymmetric constructions (with an explicit supergravity construction provided in Appendix \ref{appSUGRA}) and incorporate only generic operators in the action that would be expected in any self-consistent effective field theory treatment at high energies. 

Despite being multifield by construction, the inflationary dynamics in these models rapidly relax to effectively single-field evolution along a smooth large-field plateau in the effective potential (much as in closely related models \cite{Kaiser:2012ak,Kaiser:2013sna,Schutz:2013fua,Kaiser:2015usz}), thereby yielding predictions for primordial observables in close agreement with the latest measurements of the cosmic microwave background (CMB) radiation. Models within this family also yield efficient reheating following the end of inflation \cite{Bezrukov:2008ut,Garcia-Bellido:2008ycs,Child:2013ria,DeCross:2015uza,DeCross:2016fdz,DeCross:2016cbs,Figueroa:2016dsc,Repond:2016sol,Ema:2016dny,Sfakianakis:2018lzf,Rubio:2019ypq,Nguyen:2019kbm,vandeVis:2020qcp,Iarygina:2020dwe,Ema:2021xhq,Figueroa:2021iwm,Dux:2022kuk}. In addition, the potentials we study here include small-field features that can induce a brief phase of ultra-slow-roll evolution prior to the end of inflation, which yield sharp spikes in the power spectrum of curvature perturbations on length-scales exponentially shorter than the CMB pivot scale $k_* = 0.05 \, {\rm Mpc}^{-1}$. Upon re-entering the Hubble radius after the end of inflation, these amplified short-scale perturbations induce gravitational collapse to PBHs.

 As in previous studies of PBH formation following an ultra-slow-roll phase during inflation \cite{Garcia-Bellido:2017mdw,Ezquiaga:2017fvi,Kannike:2017bxn,Germani:2017bcs,Motohashi:2017kbs,Di:2017ndc,Ballesteros:2017fsr,Pattison:2017mbe,Passaglia:2018ixg,Byrnes:2018txb,Biagetti:2018pjj,Carrilho:2019oqg,Inomata:2021tpx,Inomata:2021uqj,Pattison:2021oen}, we find that in order to generate PBHs near the mass-range that could account for the present-day dark-matter abundance we must fine-tune one dimensionless coupling constant to several significant digits. Nonetheless, by incorporating only one fine-tuned constant, these models yield accurate predictions for eight distinct quantities---including the spectral index $n_s (k_*)$ and its running $\alpha (k_*)$, the tensor-to-scalar ratio $r (k_*)$, the isocurvature fraction $\beta_{\rm iso} (k_*, t_{\rm end})$ and primordial non-Gaussianity $f_{\rm NL}$, among others---using fewer than eight free parameters.

In future work we plan to examine the dynamics of these models across their full parameter space, including cases in which we relax the strict symmetry among the coupling constants of Eq.~(\ref{bcxisymmetries}). 
Some of these models may give rise to stochastic gravitational waves signals, which in principle could be observable with next-generation experiments~\cite{Balaji:2022dbi} such as LISA~\cite{2017arXiv170200786A,Barausse:2020rsu}, the Einstein Telescope (ET)~\cite{Maggiore:2019uih}, and DECIGO~\cite{Yagi:2011wg,Kawamura:2020pcg}.
This is an area of further research.

For each of the parameter sets we examined in this paper, quantum diffusion effects remained subdominant. However, we have found that the system's dynamics are quite sensitive to small changes in various parameters. We therefore plan to investigate regions of parameter space in which quantum effects become dominant. In such cases, the system would only be able to reach the global minimum of the potential via quantum tunnelling. For these cases, it will be important to compare the tunnelling rate to the rate of classical evolution through the ultra-slow-roll phase.

Furthermore, along the lines of recent investigations into phenomena such as the critical Higgs self-coupling \cite{Geller:2018xvz,Giudice:2021viw}, we also intend to investigate the applicability to our class of models of self-organized criticality. In particular, we are interested in the possibility that parameter sets such as those considered in Figs.~\ref{fig:VE}--\ref{fig:muomegaPR} are nearby to critical points in parameter space which act as attractors.

Other possibilities to investigate include effects on observable features of these models that arise from terms that we have thus far neglected, such as a direct quadratic coupling $b_{12} \mu \Phi_1 \Phi_2$ among the chiral superfields in the superpotential $\tilde{W}$ of Eq.~(\ref{Wtilde1}) or the addition of additional interacting fields beyond only two. (After all, the Minimal Supersymmetric Standard Model includes seven chiral superfields, each with an associated complex-valued scalar field \cite{Fayet:1976cr,Nilles:1983ge}.) In addition, we plan to investigate implications for the predicted mass distribution of PBHs produced in these models from non-Gaussianities in the probability distribution function for large-amplitude curvature perturbations.
Such modifications to the probability distribution could arise from quantum-stochastic effects during the phase of ultra-slow-roll evolution. 

%{\it Note:} As we were completing this paper, Ref.~\cite{Frolovsky:2022ewg} appeared on arXiv, which analyzes PBH formation following inflation in $F(R)$ models such as Starobinsky inflation. Following a conformal transformation, such models can be put into a (single-field) form similar to the models on which we have focused in this paper. Our findings are consistent with those reported in Ref.~\cite{Frolovsky:2022ewg}.

\section*{Acknowledgements}

We gratefully acknowledge helpful discussions with Elba Alonso-Monsalve, Alan H.~Guth, Vincent Vennin, and Shyam Balaji. Portions of this work were conducted in MIT's Center for Theoretical Physics and supported in part by the U.~S.~Department of Energy under Contract No.~DE-SC0012567. WQ was supported by the Graduate Research Fellowship Program of the U.S.~National Science Foundation. EM is supported in part by a Discovery Grant from the National Science and Engineering Research Council of Canada.

\appendix
\section{Perturbations in Multifield Models}
\label{appPerturbations}

We consider scalar perturbations around a spatially flat Friedmann-Lema\^{i}tre-Robertson-Walker (FLRW) line element,
%%%%
\beq
\begin{split}
ds^2 &= - (1 + 2A) dt^2 + 2a (t) (\partial_i B) dt dx^i \\
&\quad\quad + a^2 (t) \left[ (1 - 2 \psi) \delta_{ij} + 2 \partial_i \partial_j E \right] dx^i dx^j .
\end{split}
\label{ds}
\eeq
Gauge freedom means that only two of the four metric functions $A$, $B$, $\psi$, and $E$ in Eq.~(\ref{ds}) are independent. The field fluctuations $\delta \phi^I$ introduced in Eq.~(\ref{phivarphi}) are also gauge-dependent. We construct the gauge-invariant Mukhanov-Sasaki variables as linear combinations of field fluctuations and metric perturbations \cite{Kaiser:2012ak,Bassett:2005xm,Gong:2016qmq},
%%%%%
\beq
Q^I \equiv \delta \phi^I + \frac{ \dot{\varphi}^I }{ H} \psi ,
\label{QIdef}
\eeq
and project the perturbations $Q^I$ into adiabatic ($Q_\sigma$) and isocurvature ($Q_s$) components as in Eqs.~(\ref{QIadiso})--(\ref{deltasQs}). The equations of motion for modes $Q_\sigma (k, t)$ and $Q_s (k, t)$ then take the form \cite{Kaiser:2012ak}
%%%%%
\beq
\begin{split}
    \ddot{Q}_\sigma + 3 H \dot{Q}_\sigma + &\left[ \frac{ k^2}{a^2} + {\cal M}_{\sigma\sigma} - \omega^2 - \frac{1}{ M_{\rm pl}^2 a^3} \frac{d}{dt} \left( \frac{ a^3 \dot{\sigma}^2}{H} \right) \right] Q_\sigma \\
    &= 2 \frac{d}{dt} \left( \omega Q_s \right) - 2 \left( \frac{ V_{, \sigma}}{\dot{\sigma}} + \frac{ \dot{H}}{H} \right) \left( \omega Q_s \right) 
    \label{Qsigmaeom}
\end{split}
\eeq
and
%%%%
\beq
\ddot{Q}_s + 3 H \dot{Q}_s + \left[ \frac{k^2}{a^2} + \mu_s^2 \right] Q_s 
= 4 M_{\rm pl}^2 \frac{ \omega}{\dot{\sigma}} \frac{ k^2}{a^2} \Psi ,
\label{Qseom}
\eeq
where $\omega \equiv \epsilon^{IJ} \hat{\sigma}_I \omega_J = \pm \vert \omega^I \vert$ is the scalar turn rate \cite{Achucarro:2016fby,McDonough:2020gmn}. The gauge-invariant Bardeen potential $\Psi \equiv \psi + a^2 H (\dot{E} - B  a^{-1})$ may be related to $Q_\sigma$ and $Q_s$ via the $00$ and $0i$ components of the Einstein field equations \cite{Kaiser:2012ak}; the form of Eq.~(\ref{Qseom}) is particularly convenient for understanding the behavior of the isocurvature modes $Q_s (k, t)$ in the long-wavelength limit, $k \ll a H$. The mass matrix for the perturbations is given by
%%%%%%%%
\beq
{\cal M}^I_{\>\> J} \equiv {\cal G}^{IK} \left( {\cal D}_J {\cal D}_K V \right) - {\cal R}^I_{\>\> LMJ} \dot{\varphi}^L \dot{\varphi}^M 
\label{MIJdef}
\eeq
with the projections
%%%%%%
\beq
{\cal M}_{\sigma \sigma} \equiv \hat{\sigma}_I \hat{\sigma}^J {\cal M}^I_{\>\> J}, \>\> {\cal M}_{ss} \equiv \hat{s}_I^{\>\> J} {\cal M}^I_{\>\> J}
\label{Mprojections}
\eeq
and the mass of the isocurvature perturbations is
%%%%%
\beq
\mu_s^2 \equiv {\cal M}_{ss} + 3 \omega^2 .
\label{mus}
\eeq
In Eq.~(\ref{MIJdef}), ${\cal R}^I_{\>\> LMJ}$ is the Riemann tensor for the field-space manifold. 

When the isocurvature modes remain heavy ($\mu_s^2 \gg H^2$) and/or the turn-rate remains negligible ($\omega^2 \ll H^2$), the predictions for CMB observables revert to covariant versions of the familiar single-field forms \cite{Kaiser:2012ak,Kaiser:2013sna}. In particular, if the adiabatic perturbations remain light during inflation and we initialize the gauge-invariant perturbations in the usual Bunch-Davies vacuum state, then at Hubble crossing, solutions of Eq.~(\ref{Qsigmaeom}) will have amplitude \cite{Gordon:2000hv,Wands:2002bn,Bassett:2005xm}
%%%%%
\beq
\vert Q_\sigma (k_*, t_*) \vert = \frac{ H (t_*) }{\sqrt{ 2 k_*^3}} 
\label{Qsigmastar}
\eeq
up to an irrelevant phase, where $t_*$ is the time when $k_* = a (t_*) H (t_*)$ during inflation. Then Eqs.~(\ref{Rdef}) and (\ref{PRdef}) yield
%%%%%%%
\beq
{\cal P}_{\cal R} (k_*) = \frac{ H^2 (t_*) }{8 \pi^2 M_{\rm pl}^2 \epsilon (t_*) } .
\label{PRHepsilon}
\eeq
The spectral index $n_s (k_*)$ at some pivot scale $k_*$ is given by \cite{Kaiser:2012ak}
%%%%%
\beq
n_s (k_*) \equiv 1 + \left( \frac{d\, {\rm ln} {\cal P}_{\cal R} (k) }{ d \,{\rm ln} k} \right) \Big\vert_{k_*} \simeq 1 - 6 \epsilon (t_*) + 2 \eta (t_*) 
\label{nsdef}
\eeq
to first order in slow-roll parameters, where $\epsilon (t)$ and $\eta (t)$ are defined in Eqs.~(\ref{epsilon}) and (\ref{etadef}). 
The expression for $n_s (k_*)$ in Eq.~(\ref{nsdef}) is easiest to derive by using the usual slow-roll relation $(dx / d {\rm ln} k )\vert_{k_*} \simeq \dot{x} / H (t_*)$ at Hubble crossing \cite{Bassett:2005xm}. Likewise, the running of the spectral index is given by
%%%%
\beq
\alpha (k_*) \equiv \left( \frac{ d n_s (k) }{ d \, {\rm ln} k} \right) \Big\vert_{k_*} \simeq \left( \frac{ \dot{n}_s (k)}{H} \right) \Big \vert_{k_*}.
\label{alphadef}
\eeq
The tensor-to-scalar ratio is given by \cite{Kaiser:2013sna,Bassett:2005xm,Gong:2016qmq}
%%%%%
\beq
r (k_*) = 16 \epsilon (t_*) .
\label{rTtoS}
\eeq

For multifield models, we may compare the power spectra of curvature and isocurvature perturbations. If we adopt the conventional normalization \cite{Kaiser:2012ak,Gordon:2000hv,Bassett:2005xm,Gong:2016qmq}
%%%%
\beq
{\cal S} \equiv \frac{ Q_s }{M_{\rm pl} \sqrt{2 \epsilon}},
\label{calSdef}
\eeq
then the dimensionless isocurvature power spectrum may be written
%%%%%
\beq
{\cal P}_{\cal S} (k) \equiv \frac{ k^3}{2 \pi^2} \vert {\cal S}_k \vert^2 .
\label{PSdef}
\eeq
The isocurvature fraction $\beta_{\rm iso} (k_*, t)$ is defined as
%%%%
%%%%%
\beq
\beta_{\rm iso} (k_*, t) \equiv \frac{ {\cal P}_{\cal S} (k_*, t) }{ \left[ {\cal P}_{\cal R} (k_*, t) + {\cal P}_{\cal S} (k_*, t) \right]} .
\label{betaisodef}
\eeq
For inflationary trajectories along which the isocurvature modes remain heavy, $\mu_s^2 \gg H^2$ (as in Fig.~\ref{fig:muomegaPR}), the amplitude of isocurvature perturbations falls as $Q_s (k_*, t) \simeq Q_s (k_* , t_*) [ a (t_*) / a (t)]^{3/2}$ for times $t > t_*$. If $\vert Q_s (k_*, t_*) \vert = H (t_*) / \sqrt{ 2 k_*^3}$, akin to Eq.~(\ref{Qsigmastar}), then the amplitude of the mode ${\cal S} (k_*, t)$ will evolve for times $t > t_*$ as 
%%%
\beq
\vert {\cal S} (k_*, t) \vert \simeq \frac{ H (t_*) e^{- 3 (N_* - N(t))/2 }}{ 2 M_{\rm pl} \sqrt{ k_*^3 \, \epsilon (t) }} ,
\label{Skstar}
\eeq
where $N(t) \leq N_*$ is the number of efolds before the end of inflation. Then 
%%%%
\beq
{\cal P}_{\cal S} (k_*, t) \simeq \frac{ H^2 (t_*)}{8 \pi^2 M_{\rm pl}^2 \epsilon (t) } e^{-3 (N_* - N (t))} .
\label{PSheavy}
\eeq
Meanwhile, for $\omega^2 \ll H^2$, the amplitude of the mode ${\cal R} (k_*, t)$ remains frozen for $t > t_*$, so ${\cal P}_{\cal R} (k_*, t) = {\cal P}_{\cal R} (k_*, t_*)$, with magnitude given in Eq.~(\ref{PRHepsilon}). In that case, ${\cal P}_{\cal S} (k_*, t) \ll {\cal P}_{\cal R} (k_*, t)$ for $t > t_*$, and we find
%%%%%
\beq
\beta_{\rm iso} (k_*, t) \simeq \frac{  \epsilon (t_*) }{ \epsilon (t) } e^{-3 (N_* - N(t))} .
\label{betaisoheavy}
\eeq
For $\mu_s^2 \gg H^2$ and $\omega^2 \ll H^2$, the isocurvature fraction is therefore exponentially suppressed by the end of inflation, $\beta_{\rm iso} (k_*, t_{\rm end}) \simeq \epsilon (t_*) e^{-3 N_*} \ll 1$  \cite{Schutz:2013fua,Gordon:2000hv,Wands:2002bn,Bassett:2005xm,DiMarco:2002eb,Peterson:2010np}. 

Similarly, for heavy isocurvature modes ($\mu_s^2 \gg H^2$) and weak turning ($\omega^2 \ll H^2$), the non-Gaussianity also behaves much as in single-field models. In particular, for multifield models with curved field-space manifolds, the dimensionless coefficient $f_{\rm NL}$ may be written \cite{Seery:2005gb,Langlois:2008mn,Gong:2011cd,Elliston:2012ab,Kaiser:2012ak}
%%%%%%
\beq
\begin{split}
f_{\rm NL} &= - \frac{5}{6} \frac{ N^{, A} N^{, B} {\cal D}_A {\cal D}_B N}{ (N_{, I} N^{, I} )^2} \\
&\quad\quad - \frac{5}{6} \frac{ N_{, A} N_{, B} N_{, C} \, {\cal A}^{ABC} (k_1, k_2, k_3)}{(N_{, I} N^{, I} )^2 \sum k_i^2 },
\end{split}
\label{fNLdef}
\eeq
where $N = {\rm ln} \, a (t_{\rm end}) - {\rm ln} \, a (t_*)$ is the number of efolds before the end of inflation when the mode with comoving wavenumber $k_*$ first crossed outside the Hubble radius. The term ${\cal A}^{ABC} (k_i)$ vanishes for flat field-space manifolds, ${\cal G}_{IJ} = \delta_{IJ}$; for the curved field-space manifold we consider here, most contributions to ${\cal A}^{ABC}$ vanish identically for equilateral configurations ($k_1 = k_2 = k_3 = k_*$), and (for arbitrary shape functions) the terms proportional to ${\cal A}^{ABC}$ remain subdominant to the contributions arising from the first term in Eq.~(\ref{fNLdef}) \cite{Kaiser:2012ak}. In addition, if the isocurvature modes remain heavy during inflation, then the dominant contribution to the bispectrum arises from variations of $N$ due to fluctuations along the fields' direction of motion. In that case, Eq.~(\ref{fNLdef}) reduces to
%%%%%%
\beq
f_{\rm NL} \simeq - \frac{5}{6} \frac{ \hat{\sigma}^A \hat{\sigma}^B {\cal D}_A {\cal D}_B N }{ (\hat{\sigma}^I {\cal D}_I N )^2 } .
\label{fNLSFA1}
\eeq
Recall that $\dot{\varphi}^I {\cal D}_I A^J = {\cal D}_t A^J$ is the covariant directional derivative of vector $A^J$ in the field space. Hence for the term in the denominator of Eq.~(\ref{fNLSFA1}), we may write
%%%%%
\beq
\hat{\sigma}^I {\cal D}_I N = \frac{1}{ \dot{\sigma} } {\cal D}_t N = - \frac{ H}{\dot{\sigma} } .
\label{fNLdenom}
\eeq
For the numerator of Eq.~(\ref{fNLSFA1}), we may write
%%%%%
\beq
\hat{\sigma}^A \hat{\sigma}^B {\cal D}_A {\cal D}_B N = \hat{\sigma}^A {\cal D}_A \hat{\sigma}^B {\cal D}_B N - \frac{1}{\dot{\sigma}}\omega^B {\cal D}_B N ,
\label{fNLnum1}
\eeq
upon using the definition of the turn-rate vector $\omega^I$ in Eq.~(\ref{omegadef}). We note that 
%%%%
\beq
{\cal D}_B N = -\frac{ H}{\dot{\varphi}^B} = -\frac{ H \hat{\sigma}_B}{\dot{\sigma}} ,
\eeq
and hence the term proportional to $\omega^B$ in Eq.~(\ref{fNLnum1}) vanishes, given the orthogonality of $\omega^B$ and $\hat{\sigma}^B$. Again using $\dot{\varphi}^I {\cal D}_I A^J = {\cal D}_t A^J$, we then have
%%%%%%
\beq
\begin{split}
\hat{\sigma}^A {\cal D}_A \hat{\sigma}^B {\cal D}_B N &= \left( \frac{ H^2}{ \dot{\sigma}^2} \right) \left[ - \frac{  \dot{H}}{H^2} + \frac{ \ddot{\sigma}}{H \dot{\sigma}} \right] \\
&=  \left( \frac{ H^2}{ \dot{\sigma}^2} \right) \left( 2 \epsilon - \eta  \right) ,
\end{split}
\label{fNLnum2}
\eeq
upon using the definitions of $\epsilon$ in Eq.~(\ref{epsilon}), $\eta$ in Eq.~(\ref{etadef}), and the relationship in Eq.~(\ref{ddotsigmaeta}). Combining Eqs.~(\ref{fNLdenom})--(\ref{fNLnum2}), we then find for Eq.~(\ref{fNLSFA1})
%%%%%
\beq
f_{\rm NL} \simeq \frac{5}{6} \left( \eta - 2 \epsilon \right) + {\cal O} \left(\frac{ \omega^2 }{ H^2} \right) + {\cal O} \left( \frac{ H^2}{\mu_s^2} \right).
\label{fNLSFA2}
\eeq
For ordinary slow-roll evolution within a single-field attractor, we therefore find that the coefficients for equilateral, orthogonal, and local configurations of the bispectrum will each generically remain small, $\vert f_{\rm NL} \vert \lesssim {\cal O} (10^{-2})$. During ultra-slow-roll, when $\eta \rightarrow 3$, the non-Gaussianity will rise to be ${\cal O} (1)$ \cite{Kaiser:2012ak,Kaiser:2013sna,Kaiser:2015usz,Langlois:2008mn,Bernardeau:2002jy,Seery:2005gb,Yokoyama:2007dw,Byrnes:2008wi,Peterson:2010mv,Chen:2010xka,Byrnes:2010em,Gong:2011cd,Elliston:2011dr,Elliston:2012ab,Seery:2012vj,Mazumdar:2012jj,Peterson:2010np,Gong:2011uw,Gong:2016qmq}.

The comoving CMB pivot scale $k_* = 0.05 \, {\rm Mpc}^{-1}$ first crossed outside the Hubble radius $N_* \equiv N (k_*)$ efolds before the end of inflation \cite{Dodelson:2003vq,Liddle:2003as}
%%%%%
\beq
\begin{split}
N_* &= 67 - {\rm ln} \left( \frac{ k_*}{a_0 H_0} \right) + \frac{1}{4} {\rm ln} \left( \frac{ V^2 (t_*)}{ M_{\rm pl}^4 \rho (t_{\rm end}) } \right) \\
&\quad\quad + \frac{1 - 3 w_{\rm eff} }{12 (1 + w_{\rm eff} ) } {\rm ln} \left( \frac{ \rho (t_{\rm rd}) }{\rho (t_{\rm end} ) } \right) \\
&\simeq 62 + \frac{1}{4} {\rm ln} \left( \frac{ V^2 (t_*)}{3 M_{\rm pl}^6 H^2 (t_{\rm end}) } \right) - \frac{ N_{\rm reh}}{4} ,
\end{split}
\label{Nstar}
\eeq
where the subscript $0$ denotes present-day values, $t_*$ is the time when $k_* = a (t_*) H (t_*)$ during inflation, $t_{\rm end}$ is the time at which inflation ends, and $t_{\rm rd}$ is the time when the universe first attains a radiation-dominated equation of state after the end of inflation. In the second line, we assume that the reheating epoch persists for $N_{\rm reh}$ efolds after the end of inflation, during which the universe expands with a matter-dominated equation of state $w_{\rm eff} \simeq 0$ \cite{Amin:2014eta,Allahverdi:2020bys}.

\section{Realization in Supergravity}
\label{appSUGRA}

For a textbook review of supergravity, we refer the reader to Ref.~\cite{Freedman:2012zz}. For a concise review, we refer the reader to the appendices of Ref.~\cite{Kolb:2021xfn}.

The potential in Eq.~\eqref{Vtildertheta} is realized within the framework of $\mathcal{N}=1$ supergravity in $d=4$ dimensions. We take two chiral superfields $\Phi^{I}$, with $I=\{1,2\}$, with field content 
\begin{align}
    \Phi^{I}(x,\theta)=\varpi^{I}+\sqrt{2}\theta\eta^{I}+\theta\theta F^{I} ,
\end{align}
where each $\varpi^I$ (for $I \in \{ 1, 2 \}$) is a complex scalar field, each $\eta^{I}$ is a two-component Weyl spinor, $\theta$ is the fermionic coordinate on superspace, and $F^{I}$ are non-dynamical auxiliary fields; $\bar{\Phi}^{\bar{I}}$ denotes the corresponding anti-chiral superfields. Each complex scalar field $\varpi^I$ can be written in terms of its real and imaginary parts as 
\begin{align}
    \varpi^{I}=\frac{1}{\sqrt{2}}(\phi^{I}+i\psi^{I}) .
\end{align}
Our model is specified in the Jordan frame by a superpotential  $\tilde{W}(\Phi^I)$ and  K\"{a}hler potential $\tilde{K}(\Phi^I,\bar{\Phi}^{\bar{I}})$. The kinetic terms of the scalar components are given by
\begin{equation}
    \mathcal{L}_{\rm kinetic} = - \tilde{\mathcal{G}}_{I \bar{J}} \tilde{g}^{\mu \nu } \partial_{\mu} \varpi^{I} \partial_{\nu} \bar{\varpi}^{\bar{J}} ,
\end{equation}  
with field-space metric 
\begin{align}
    \tilde{\mathcal{G}}_{I\bar{J}}=\frac{\partial}{\partial\Phi^{I}}\frac{\partial}{\partial\bar{\Phi}^{\bar{J}}}\tilde{K}(\Phi^J,\bar{\Phi}^{\bar{J}})_{\Phi^I\rightarrow \varpi^I, \bar{\Phi}^{\bar{I}} \rightarrow \bar{\varpi}^{\bar{I}} }.
    \label{GtildeIJ}
\end{align}
The scalar potential in the Jordan frame is given by
\begin{equation}
     \tilde{V} = \bigg\{ e^{\tilde{K}/M_{\rm pl}^2}\left[ |D\tilde{W} |^2 - 3 M_{\rm pl}^{-2}|\tilde{W} |^2\right]\bigg\}_{\Phi^I\rightarrow \varpi^I, \bar{\Phi}^{\bar{I}} \rightarrow \bar{\varpi}^{\bar{I}} } ,
\end{equation}
where $D_I \equiv \partial_I + M_{\rm pl}^{-2} \tilde{K}_{, I}$.

We select the K\"{a}hler potential to be
\begin{equation}
    \tilde{K} = -\frac{1}{2} \displaystyle \sum _{I=1} ^2 ( \Phi^I - \bar{\Phi}^{\bar{I} })^2
    \label{Kahler1}
\end{equation}
and work with the generic superpotential
\begin{equation}
    \tilde{W}=  \sqrt{2} \mu b_{IJ} \Phi^I \Phi^J + 2 c_{IJK} \Phi^I \Phi^J \Phi^K ,
\end{equation}
where $\mu$ is a mass-scale. Given Eqs.~(\ref{GtildeIJ}) and (\ref{Kahler1}), the field-space metric in the Jordan frame is flat,
%%%%%
\beq
\tilde{\cal G}_{I \bar{J}} = \delta_{I \bar{J}} .
\label{Gtildeflat}
\eeq

For $\tilde{K}$ given in Eq.~(\ref{Kahler1}), we find $\tilde{K}\rightarrow  \sum_I (\psi^I)^2$ upon projecting $\{ \Phi^I, \bar{\Phi}^{\bar{I}} \} \rightarrow \{ \varpi^I, \bar{\varpi}^{\bar{I}} \}$; hence the imaginary components $\psi^I$ of each scalar field $\varpi^I$ become heavy, due to the exponential dependence of $\tilde{V}$ on the K\"{a}hler potential. In particular, it is straightforward to show that $m_\psi^2 \simeq 10 H^2$ (in the Einstein frame), which allows us to integrate out the imaginary components $\psi^I$ during inflation. The resulting scalar potential for the real components $\varpi^1 \equiv \phi / \sqrt{2}$ and $\varpi^2 \equiv \chi / \sqrt{2}$ is given by
\begin{widetext}
\begin{equation}
    \begin{split}
    \tilde{V}(\phi,\chi) = & \> 4 b_1^2 \mu ^2 \phi ^2-\frac{3 b_1^2 \mu ^2 \phi ^4}{2 M_{\rm pl}^2} 
     -\frac{3 b_1 b_2 \mu ^2 \chi ^2 \phi ^2}{M_{\rm pl}^2}+12 b_1 c_1 \mu  \phi ^3-\frac{3 b_1 c_1 \mu  \phi ^5}{M_{\rm pl}^2}  \\
    &+8 b_1 c_2 \mu  \chi  \phi ^2-\frac{3 b_1 c_2 \mu  \chi  \phi ^4}{M_{\rm pl}^2}+4 b_1 c_3 \mu  \chi ^2 \phi  
     -\frac{3 b_1 c_3 \mu  \chi ^2 \phi ^3}{M_{\rm pl}^2}-\frac{3 b_1 c_4 \mu  \chi ^3 \phi ^2}{M_{\rm pl}^2}+4 b_2^2 \mu ^2 \chi ^2  \\
    & -\frac{3 b_2^2 \mu ^2 \chi ^4}{2 M_{\rm pl}^2}-\frac{3 b_2 c_1 \mu  \chi ^2 \phi ^3}{M_{\rm pl}^2}+4 b_2 c_2 \mu  \chi  \phi ^2  
     -\frac{3 b_2 c_2 \mu  \chi ^3 \phi ^2}{M_{\rm pl}^2}+8 b_2 c_3 \mu  \chi ^2 \phi -\frac{3 b_2 c_3 \mu  \chi ^4 \phi }{M_{\rm pl}^2} \\
    & +12 b_2 c_4 \mu  \chi ^3-\frac{3 b_2 c_4 \mu  \chi ^5}{M_{\rm pl}^2}-\frac{3 c_1^2 \phi ^6}{2 M_{\rm pl}^2}+9 c_1^2 \phi ^4  
    -\frac{3 c_1 c_2 \chi  \phi ^5}{M_{\rm pl}^2}+12 c_1 c_2 \chi  \phi ^3-\frac{3 c_1 c_3 \chi ^2 \phi ^4}{M_{\rm pl}^2}  \\
    & +6 c_1 c_3 \chi ^2 \phi ^2 -\frac{3 c_1 c_4 \chi ^3 \phi ^3}{M_{\rm pl}^2}-\frac{3 c_2^2 \chi ^2 \phi ^4}{2 M_{\rm pl}^2} 
    +4 c_2^2 \chi ^2 \phi ^2+c_2^2 \phi ^4-\frac{3 c_2 c_3 \chi ^3 \phi ^3}{M_{\rm pl}^2}  \\
    & +4 c_2 c_3 \chi  \phi ^3+4 c_2 c_3 \chi ^3 \phi -\frac{3 c_2 c_4 \chi ^4 \phi ^2}{M_{\rm pl}^2}  
     +6 c_2 c_4 \chi ^2 \phi ^2-\frac{3 c_3^2 \chi ^4 \phi ^2}{2 M_{\rm pl}^2}+c_3^2 \chi ^4+4 c_3^2 \chi ^2 \phi ^2 \\
    & -\frac{3 c_3 c_4 \chi ^5 \phi }{M_{\rm pl}^2}+12 c_3 c_4 \chi ^3 \phi -\frac{3 c_4^2 \chi ^6}{2 M_{\rm pl}^2}+9 c_4^2 \chi ^4  ,
\end{split}
\label{Vtildefull}
\end{equation}
\end{widetext}
where, as noted below Eq.~(\ref{Wtilde}), we define $b_1 \equiv b_{11}$, $b_2 \equiv b_{22}$, $c_1 \equiv c_{111}$, $c_2 \equiv (c_{112}+c_{121}+c_{211})$, $c_3 \equiv (c_{122}+c_{212}+c_{221})$, and $c_4 \equiv c_{222}$. If one considers inflationary models with $\xi \gg 1$, the perturbation modes accessible to observation correspond to the those that exited the Hubble radius when $\phi,\chi \ll M_{\rm pl}$. Taking the $\phi,\chi \ll M_{\rm pl}$ limit, Eq.~(\ref{Vtildefull}) simplifies to
\begin{eqnarray}
    \tilde{V}(\phi,\chi) =&& 4 b_1^2 \mu ^2 \phi ^2+12 b_1 c_1 \mu  \phi ^3+8 b_1 c_2 \mu  \chi  \phi ^2 \\
    && +4 b_1 c_3 \mu  \chi ^2 \phi +4 b_2^2 \mu ^2 \chi ^2+4 b_2 c_2 \mu  \chi  \phi ^2 \nonumber \\
    && +8 b_2 c_3 \mu  \chi ^2 \phi +12 b_2 c_4 \mu  \chi ^3+9 c_1^2 \phi ^4+12 c_1 c_2 \chi  \phi ^3 \nonumber \\
    && +6 c_1 c_3 \chi ^2 \phi ^2+4 c_2^2 \chi ^2 \phi ^2+c_2^2 \phi ^4+4 c_2 c_3 \chi  \phi ^3 \nonumber \\
    && +4 c_2 c_3 \chi ^3 \phi +6 c_2 c_4 \chi ^2 \phi ^2+c_3^2 \chi ^4+4 c_3^2 \chi ^2 \phi ^2 \nonumber \\
    && +12 c_3 c_4 \chi ^3 \phi +9 c_4^2 \chi ^4 + \mathcal{O}\left(\phi^5/M_{\rm pl}, \chi^5/M_{\rm pl} \right) \nonumber .
\end{eqnarray}
We note that the benchmark value of $\xi$ in Higgs inflation is ${\cal O}(10^4)$ \cite{Bezrukov:2007ep}, and further note that our model can accommodate $\xi$ over many orders of magnitude, via the rescaling of Eq.~\eqref{bcscaling2}. Finally, translating to polar coordinates, we arrive at Eq.~\eqref{Vtildertheta}.

These models can easily be unified with the current epoch of cosmic acceleration and the observed cosmological constant. This is done by introducing an additional superfield $S$ which satisfies a nilpotency constraint,
\begin{equation}
    S(x,\theta)^2  =0 .
\end{equation}
This condition projects out the scalar component of $S$ from the bosonic sector of the theory. The cosmological applications of the nilpotent superfields were developed in, e.g., Refs.~\cite{Ferrara:2014kva,McDonough:2016der,Kallosh:2017wnt}. The simplest model is given by, 
\begin{equation}
    W = M S   \,\, , \,\, K = S \bar{S},
\end{equation}
leading to a scalar potential which is simply a cosmological constant
\begin{equation}
    V  = M^2 .
\end{equation}
Inflation and dark energy can be realized in this context either by promoting $M$ to a function of fields, or else through field-dependent corrections to the K\"{a}hler potential such as \cite{McDonough:2016der},
\begin{equation}
  \delta K =  f(\Phi,\bar{\Phi}) S \bar{S}  .
\end{equation}
In both cases the scalar potential is simply,
\begin{equation}
    V =  G^{S \bar{S}} \partial_{S}W \partial_{\bar{S}}\bar{W}  .
\end{equation}

We may easily combine the nilpotent superfield models with the inflation models proposed in this paper. For example, we may consider,
\begin{equation}
\begin{split}
     \tilde{W} &=  M S + \tilde{W}_{\rm infl}(\Phi^I) , \\ 
     \tilde{K} &= S \bar{S} + \tilde{K}_{\rm infl}(\Phi^I,\bar{\Phi}^{\bar{I}}) ,
\end{split}
\end{equation}
where $\tilde{W}_{\rm infl}$ and $\tilde{K}_{\rm infl}$ refer to the Jordan-frame $W$ and $K$ of our multifield inflation model. The resulting (Jordan-frame) scalar potential is given by,
\begin{equation}
    \tilde{V} = M^2 + \tilde{V}_{\rm infl}(\phi,\chi) ,
\end{equation}
where $\tilde{V}_{\rm infl}$ is the Jordan frame inflationary potential of our two-field model. This approach allows for additional spectator fields during inflation, simply by promoting $M$ to a function of fields, or by corrections to $\tilde{K}$ \cite{McDonough:2016der}.

Finally, nonminimal couplings of the superfields $\Phi^I$ to gravity, in a manifestly supersymmetric form, can be accomplished following the procedure of Ref.~\cite{Kallosh_2013}, slightly generalized from one inflaton to two.

\section{Analytic Solution for the Background Fields' Trajectory}
\label{appTheta}

As noted in Section \ref{sec:trajectories}, if the dimensionless couplings obey the symmetries of Eq.~(\ref{bcxisymmetries}), then we may solve analytically for the background fields' trajectory during inflation. We identify local minima of the potential in the angular direction by calculating
%%%%%
\beq
\begin{split}
    V_{, \theta} (r, \theta) &= \frac{ M_{\rm pl}^4}{[2 f (r) ]^2} \left[ {\cal C}' (\theta) \mu r^3 + {\cal D}' (\theta) r^4 \right] \\
    &= F( r) \, G (r, \theta) ,
\end{split}
\label{Vcommatheta1}
\eeq
where $F (r)$ is some function independent of $\theta$, and
%%%%%
\beq
G (r, \theta) \equiv {\cal C}' (\theta) \mu + {\cal D}' (\theta) r .
\label{Grthetadef}
\eeq
The system will evolve along local minima $\theta_*$ such that $V_{, \theta} (r, \theta_*) = 0$, which corresponds to $G(r, \theta_*)$ = 0. Given the definitions of ${\cal C} (\theta)$ and ${\cal D} (\theta)$ in Eq.~(\ref{BCDdef}), the terms that appear in $G (r, \theta)$ may be written
%%%%%
\beq
\begin{split}
    {\cal C}' (\theta) &= - 18 b c_1 \sin (2 \theta) \left[ \cos \theta - \left( \frac{ c_4 }{c_1} \right) \sin \theta \right] + 12 b c_2 g_1 (\theta) , \\
    {\cal D}' (\theta) &= - 18 c_1^2 \sin (2\theta) \left[ \cos^2 \theta - \left( \frac{c_4 }{c_1} \right)^2 \sin^2 \theta \right] + 4 c_2 g_2 (\theta)
\end{split}
\label{CprimeDprime}
\eeq
with
%%%%%
\beq
\begin{split}
    g_1 (\theta) &\equiv \cos^3 \theta + \sin (2\theta) \left( \cos \theta - \sin \theta \right) - \sin^3 \theta , \\
    g_2 (\theta) &\equiv (3 c_1 + c_2 ) \cos^4 \theta \\
    &\quad + \frac{3}{2} (c_1 + c_2 + c_4) \sin (2 \theta) \left( \cos^2 \theta - \sin^2 \theta \right) \\
    &\quad - 9 (c_1 - c_4) \cos^2 \theta \sin^2 \theta - (3 c_4 + c_2 ) \sin^4 \theta.
\end{split}
\label{g1g2}
\eeq
Closed-form solutions to the equation $G (r, \theta_*) = 0$ may then be found by using the substitution $\theta_* (r) = {\rm arccos} (x (r))$, resulting in the expression for $x^\pm (r)$ given in Eq.~(\ref{xpm}).

%\bibliography{PBH.bib}


\end{document}